\documentclass[structabstract]{aa}
\usepackage{graphicx}
\usepackage{txfonts}
\usepackage{longtable}

\usepackage{enumerate}
\usepackage{booktabs}

\begin{document}
   \title{Simultaneous formation of Solar System giant planets}

   \subtitle{}

   \author{
          O. M. Guilera
          \inst{1,2}\fnmsep\thanks{Fellow of the Consejo Nacional de
	  Investigaciones  Cient\'{\i}ficas y T\'ecnicas (CONICET), Argentina.
          E-mail: oguilera@fcaglp.unlp.edu.ar}
          \and
          A. Fortier\inst{3}\fnmsep\thanks{   
	  E-mail: andrea.fortier@space.unibe.ch}
	  \and	
          A. Brunini\inst{1,2}\fnmsep\thanks{Member of the Carrera del
Investigador Cient\'{\i}fico, CONICET, Argentina.  
	  E-mail: abrunini@fcaglp.unlp.edu.ar}
          \and
          O. G. Benvenuto
	  \inst{1,2}\fnmsep\thanks{Member of the Carrera del 
	  Investigador Cient\'{\i}fico, Comisi\'on de Investigaciones 
	  Cient\'{\i}ficas de la Provincia de Buenos Aires, Argentina.
	  E-mail: obenvenu@fcaglp.unlp.edu.ar}
          }
   \offprints{O. M. Guilera}

   \institute{Facultad de Ciencias Astron\'omicas y Geof\'{\i}sicas, 
              Universidad Nacional de La Plata, Paseo del Bosque s/n (B1900FWA)
La Plata, 
              Argentina
              \and
	      Instituto de Astrof\'{\i}sica de La Plata, IALP,
CCT-CONICET-UNLP, 
	      Argentina
              \and
              Physikalisches Institut, University of Bern,
Sidlerstrasse 5, CH - 3012, Bern, Switzerland.\\
              }

   \date{Received; accepted}

  \abstract
   {In the last few years, the so--called ``Nice model'' has got a
   significant importance in the study of the formation and evolution of
   the solar system. According to this model, the initial orbital
   configuration of the giant planets was much more compact than the one
   we observe today.} 
   {We study the formation of the giant planets in connection with some parameters that describe the protoplanetary disk. The aim of this
   study is to establish the conditions that favor their simultaneous
   formation in line with the initial configuration proposed by the Nice
   model. We focus in the conditions that lead to the simultaneous formation of two massive cores, corresponding to Jupiter and Saturn, able to achieve the cross-over mass (where the mass of the envelope of the giant planet equals the mass of the core, and gaseous runway starts) while Uranus and Neptune have to be able to grow to their current masses.} 
   {We compute the $in\;situ$ planetary formation, employing the
   numerical code introduced in our previous work, for different density
   profiles of the protoplanetary disk. Planetesimal migration is taken
   into account and planetesimals are considered to follow a size
   distribution between $r_p^{min}$ (free parameter) and $r_p^{max}=
   100$ km. The core's growth is computed according to the oligarchic
   growth regime.} 
   {The simultaneous formation of the giant planets was successfully completed for several initial conditions of the disk. We find that
   for protoplanetary disks characterized by a power law ($\Sigma
   \propto r^{-p}$), smooth surface density profiles ($p \leq 1.5$)
   favor the simultaneous formation. However, for steep slopes ($p\sim
   2$, as previously proposed by other authors) the simultaneous
   formation of the solar system giant planets is unlikely.} 
   {The simultaneous formation of the giant planets --in the context of
   the Nice model-- is favored by smooth surface density profiles. The
   formation time-scale is in agreement with the estimates of disk
   lifetimes if a significant mass of the solids accreted by the planets
   is contained in planetesimals with radii $<$ 1~km.}


   \maketitle
%

\section{Introduction}

\label{sec:intro}

The initial configuration of the Nice model (Tsiganis et al., 2005; Gomes et
al., 2005; Morbidelli et al., 2005) represents the orbital configuration of the
outer
solar system after the gas of the primordial nebula dissipated.
The model assumes that the giant planets were initially in nearly circular and
coplanar orbits. This is compatible with the work of Thommes et al. (2008), who
find that planetary system analogues to our solar system come out if the gas
giants do not undergo significant migration during their formation and remain in
nearly circular
orbits. The Nice model proposes a compact initial configuration for the
location of the giant planets; more precisely, the giant planet system is
assumed to be in the range of $\sim 5.5$~AU to $\sim 14$~AU. The gas giants,
Jupiter and Saturn, are supposed to be close to their mutual 2:1 mean motion resonance~(MMR); Jupiter
located at $\sim 5.5$~AU and Saturn  between $8 - 8.5$~AU. This is an important
condition, required to avoid the migration of both planets during the gas disk
lifetime (Masset \& Snellgrove, 2001; Morbidelli \& Crida, 2007). Regarding the
ice giants, they are
assumed to be located at $\sim 11$ and $\sim 14$~AU. 

Another important key of the Nice model is the existence of a planetesimal disk
beyond the orbit of the giant planets. The inner edge of the disk is proposed to
be located
$\sim 16$~AU, and the outer edge is fixed at $\sim 30$~AU; the total mass of the
planetesimal disk being
$\sim 35 - 40~\mathrm{M_{\oplus}}$. The planetesimals conforming this disk
gravitationally interact
with the giant planets and cause the inward migration of Jupiter and the outward
migration of Saturn to their current positions. In this process,  Jupiter and Saturn cross their mutual
2:1 mean motion resonance, leading the 
system to undergo a phase of dynamical instability. During this phase, Uranus
and Neptune migrate outward chaotically. According to the numerical simulations
of Tsiganis et al. (2005), there is a 50~\% probability that 
the two icy planets switch places in the process.

The success of the Nice model relies on the fact that it is able to explain
quantitatively the final orbits, eccentricities, and inclinations of the giant
planets (Tsiganis et al., 2005); the chaotic capture of Jupiter's Trojan
asteroids (Morbidelli et al., 2005); the origin of the Late Heavy Bombardment
(Gomes et al., 2005); the formation of the Kuiper belt (Levison et al., 2008)
and the secular architecture of the outer solar system (Morbidelli et al.,
2009). Remarkably, the Nice model does not say anything about how the initial
conditions proposed by itself could be achieved. 

More recently, Morbidelli \& Crida (2007) -following the work of Masset \&
Snellgrove (2001)- showed that under certain parameters of the gas disk,
Jupiter and Saturn did not migrate once both planets got locked in 2:3
MMR. Morbidelli \& Crida (2007) showed that this kind of no-migrating (or
slowly migrating) evolution is possible if the mass ratio of the two planets
is similar to the mass ratio of Jupiter and Saturn. Then, Morbidelli et
al. (2007) extended this study to the four giant planets of the solar system. They locked Jupiter and Saturn in 2:3 MMR, and found that the inner ice giant could be trapped in the 2:3 or 3:4 MMR with Saturn. Then, the outer ice giant could be trapped in the 3:4, 4:5, or 5:6 MMR with the inner ice giant. They showed that the resonant structure is preserved until the gas is completely dispersed. Then, only two configurations were dynamically stable for several hundred million years. Finally, they showed that the existence of a planetesimal disk beyond the outer ice giant lead to a dynamical instability, and the essential ingredients of the Nice model are preserved. Batygin \& Brown (2010) extended this work and found more possible multi-resonant initial conditions, several of which evolved -in the presence of a planetesimal disk beyond the ice giants- preserving the essential ingredients of the Nice model. While this initial conditions are not exactly the same to that proposed by the Nice model, all of them propose a compact orbital configuration for the giant planets of the solar system in a similar way to the Nice model. However, all these studies assumed that the giant planets were already formed, and did not address the question of how they formed.

Desch (2007) was the first to investigate this topic. Adopting the {\it in situ} formation and perfect accretion for the giant planets, he recalculated the ``minimum-mass solar nebula''
 considering the initial more compact configuration of the giant planets
together with
the remnant planetesimal disk, as proposed by the Nice model. He then derived a
much steeper
minimum mass solar nebula surface density profile ($\Sigma \propto a^{-2.168}$,
where $a$
is the distance to the Sun)
than the classical
one ($\Sigma \propto a^{-3/2}$, Weidenschilling, 1977; Hayashi, 1981). He also
estimated the growth time for the cores of the four giant planets. Assuming the
oligarchic growth regime for the solid embryos, relative velocities in
equilibrium, and a single size
population of planetesimals with radii of 100~m, he obtained growth times of
about 0.5~Myr for Jupiter, $1.5 - 2.0$~My for Saturn, $5.5 - 6.0$~Myr for
Neptune, and $9.5 - 10.5$~Myr for Uranus. However, he did not consider the
presence of the also growing gas envelope 
of the planets.

In a more realistic way, adopting the initial configuration for the location of
the giant planets as proposed by the Nice model (Jupiter at 5.5~AU, Saturn at
8.3~AU, Neptune at 11~AU, and Uranus at 14~AU), and employing the surface
density profile derived by Desch (2007), Benvenuto et al. (2009) calculated the
{\it in situ},
isolated formation of the giant planets of the solar system. They used the same
code
already introduced in previous works (Benvenuto \& Brunini, 2005; Fortier et
al., 2007;
2009) but incorporating a size distribution  for the planetesimals radii,
considering they take values between 30~m -
100~km (nine species geometrically evenly
spaced), with a mass
distribution $n(m) \propto m^{-\alpha}$, with $\alpha= 2.5$. The choice of
$r_p^{min}= 30$~m as the minimum radius of the size distribution of planetesimals 
relied on a simple estimation for the planetesimal radius for which the
migration time-scale of smaller planetesimals should be shorter than the
protoplanet's planetesimal accretion time-scale, so the accretion of
planetesimals with radii $< 30$~m should be not so efficient. Imposing a value
of
$11 \, \mathrm{g~cm^{-2}}$ for the solids surface density at the location of
Jupiter, they calculated the formation time for Jupiter (0.44~Myr), Saturn
(1.4~Myr), Neptune (2.5~Myr), and Uranus (4.75~Myr). Furthermore, they found that
the final core masses of the four planets were in a very good agreement with the
present theoretical estimations. However, some important simplifications  were
assumed in that work.  It is well known that planet formation, disk evolution
and planetesimal migration occur on the same time scale. Then, a closer approach
to the formation of the giant planets should include an approximation of the
other two processes. In Benvenuto et al. (2009) -and also in Desch (2007)- the migration of planetesimals
due to the gas drag of
the disk was not taken into
account. As it was shown by Thommes et al.~(2003), Chambers~(2006), and
Brunini \& Benvenuto~(2008) this
effect has a strong influence on the accretion time, especially for small
planetesimals. Then, planetesimal migration should be
incorporated in a more realistic model. On the other hand, the evolution of the
gaseous
component of the disk should be taken into account, as the gas density is
an input parameter when determining the accretion rates and the
planetesimals migration velocities. Finally, the planetesimals relative
velocities  were
considered to be in
equilibrium. Nevertheless, it is not clear that an equilibrium condition is ever
achieved, specially for small planetesimals (Chambers, 2006; Brunini \&
Benvenuto, 2008). Instead, relative
velocities out of equilibrium should be adopted due to the fact that the size
distribution of the planetesimals extends to meter-size objects.

These conditions were incorporated in a recent model developed by Guilera
et al. (2010). They calculated the simultaneous formation of Jupiter and Saturn
-at their current locations- immersed in a protoplanetary disk that evolves with
time. They found that the simultaneous formation of Jupiter and Saturn could be
very different when compared to the isolated formation of each planet. For the
classical solar system surface density profile ($\Sigma \propto a^{-3/2}$), they
found that the formation of Jupiter inhibits the growth of Saturn. On the other
hand, for a disk with surface density profile $\Sigma \propto a^{-1/2}$, Saturn
turned out to form faster than Jupiter. Moreover, the quick formation of Saturn
produced an inner
solid density wave that significantly reduced the time of the formation of
Jupiter.
Finally, for a disk with surface density profile $\Sigma \propto a^{-1}$,
simultaneous and isolated formation of both planets resulted in  very similar
outcomes.

The aim of this work is to study some of the parameters that characterize the protoplanetary disk in order to determine the conditions that could favor/not favor the formation of the solar system giant planets, in line with the initial configuration proposed by the Nice model. The formation of the planets is calculated using the {\it in situ} approximation. However, the formation of the four planets are computed simultaneously, which means that the formation of one planet can affect the formation of another one because it perturbs the planetesimal disk. We will focus then our analysis on the first stage of the planetary formation, this meaning, before the gaseous runaway starts. We aim to have an insight into the conditions of the disk (especially, the surface density profiles and the planetesimal size distribution) that lead to the formation of the cores of Jupiter and Saturn on a timescale that enable the planets to reach the gaseous runaway. Moreover, Uranus and Neptune should also be able to grow to their current mass. For this purpose, we consider a population of planetesimals with
non homogeneous radii following a power law distribution, where the radii go from $r_p^{min}$ to 100~km (the maximum value of the distribution is fixed for all the cases). The value of the minimum radius is to be determined as follows: we first calculate the isolated formation of the four planets for different values of $r_p^{min}$ and determine then the optimum values they can take (by ``optimum values'' we mean those values of $r_p^{min}$ that the four planets have in common 
that minimize their time of formation ). With these values we then explore the simultaneous formation. This procedure is employed to analyze  different profiles for the surface density of the protoplanetary disk. Finally, we discuss the scenario that best matches the basic observational constraints for the formation of the giant planets (the disk dissipation time scale and the estimated mass of the core) with the conditions of the Nice model.

The reminder of this paper is organized as follows: In Section 2 we give a brief
description of our model, developed in a previous work (Guilera et al., 2010).
In Section 3 we calculate the formation of the giant planets of the solar system
for a steep surface density profile similar to the one derived by Desch (2007).
Then, in Section 4 we present the results for smoother surface density profiles.
Finally, in Section 5 we discuss our results and present the concluding
remarks.


\section{A brief description of the model}

\label{sec:sec2}  

In a previous work (Guilera et al., 2010), we developed a numerical code to
compute {\it in situ} but simultaneous formation of several giant planets immersed in a protoplanetary disk that
evolves with time. In the present paper, we used this
code to study the first stage of the formation of the four giant
planets of the solar system. 

The protoplanetary disk was characterized by gaseous and solid (a planetesimal disk) components. We considered a classic power-law to describe the surface density,
$\Sigma \propto a^{-p}$. In our model, the disk extends from 0.4 to 30~AU.
We fixed the outer edge of the disk at 30~AU in order to be consistent
with the outer edge of the planetesimal disk used by the Nice model. For the planetesimal disk, we considered a population of non-equal sized bodies. Planetesimals were assumed to follow a size distribution whose radii were between a minimum value, $r_p^{min}$, and 100~km, with steps selected in order that the quotient of masses of consecutive sizes is a factor of two. For example, the continuous planetesimal size distribution between 1~km and 100~km was characterized by 21 different sizes of planetesimals, while the continuous planetesimal size distribution between 10~m and 100~km was characterized by 41 different sizes of planetesimals. Kokubo \& Ida (2000), employing N-body simulations, and more recently, Ormel et al. (2010), employing statistical simulations, showed that in the oligarchic growth regime planetesimals follow a mass distribution $dn/dm \propto m^{-p}$, with $p \sim 2.5$. For this work, we adopted this distribution. Note that most of the mass is in small planetesimals.

As we mentioned above, we considered the protoplanetary disk to
evolve. In terms of the planetesimal disk, we considered planetesimal
migration due to nebular gas drag. We adopted that the migration velocities are given by the prescription of Adachi et al. (1976),
\begin{eqnarray}
\frac{da}{dt} = &-& \frac{2a}{T_{fric}}\bigg(\eta^2 + \frac{5}{8}e^2 +
 \frac{1}{2} i^2\bigg)^{1/2} \bigg[\eta + \bigg(\frac{5}{16} + \frac{\alpha}{4}\bigg)e^2 +
\frac{1}{4}i^2\bigg],
\label{eq:vel-migra}
\end{eqnarray}
where 
\begin{eqnarray}
T_{fric}= \frac{8\rho_p r_p}{3C_D\rho_{g}v_k}.
\end{eqnarray}
Here $\rho_p$ is the bulk density of planetesimals ($1.5~\mathrm{g~cm^{-2}}$), $r_p$ is the radius of the planetesimal, $C_D \sim 1$, $\eta=(v_k - v_{gas})/v_k$, and $\alpha$ is the exponent of the power-law density of the gas in the disk mid plane ($\rho_g \propto a^{-\alpha}$). Then, the evolution of the planetesimals disk obeys a continuity equation, 
\begin{eqnarray}
\frac{\partial \Sigma_s}{\partial t} - \frac{1}{a}\frac{\partial}{\partial a}
 \bigg(a\frac{da}{dt}\Sigma_s\bigg) = F(a), \label{eq:continuity}
\end{eqnarray}
where $\Sigma_s$ is the surface density of solids and $F(a)$ describes the sinks of disk material (accretion by the forming planets).

Adopting a temperature profile for an optically thin disk ($T \propto a^{-1/2}$) the density distribution of gas on the mid-plane of the disk is given by $\rho_g \propto a^{-p-5/4}$, and for simplicity, we assumed that the gaseous component dissipates out following an exponential decay of its density with a characteristic time-scale of 6~Myr (Haisch et al., 2001),
\begin{eqnarray}
\rho_g(a,t)= \rho_g(a,0)e^{-t/6~\mathrm{Myr.}}.
\end{eqnarray}

For the growth of the core, we adopted the oligarchic regime, in accordance to our previous work (Guilera et al., 2010). Adopting the {\it particle in a box} approximation (Inaba et al., 2001) for the accretion of solids
\begin{eqnarray}
\frac{dM_C}{dt}= \frac{2\pi \Sigma_s(a_P)R_H^2}{P}P_{coll},
\end{eqnarray}
and integrating over the planetesimals size distribution and the feeding zone, the solid accretion rate is given by 
\begin{eqnarray}
\frac{dM_C}{dt}= \int_{PSD} dm \int_{FZ} && 2\pi \psi(a, R_H, a_P) ~ \times \nonumber \\
&& \frac{2\pi\Sigma_s(a,m)R_H^2}{P}~P_{coll}(a,m)~a~da.  
\label{eq:tas}
\end{eqnarray}
where $M_{C}$ is the mass of the core, $R_H$ is the Hill radius, $P$ is the orbital period, $a_P$ is the semi-major axis of the planet, and $\psi$ is a normalization function (see Brunini \& Benvenuto, 2008). $P_{coll}$ is the collision probability between the planetesimals and the planet, which is a function of the planet's core radius, the Hill radius of the planet, and the relative velocity of planetesimals $P_{coll}=P_{coll}(R_C,R_H, v_{rel})$ (for more details, see Guilera et al., 2010).

We also took into account the enhancement of the planet's capture cross-section due to the presence of the planet's envelope. Inaba \& Ikoma~(2003) found an approximate solution to the equations of motion, which allows for a rapid estimation of the radius of the planetesimal captured, $r_p$, as function of the the planet's enhanced radius $\tilde{R}_C$
\begin{equation}
 r_p= \frac{3 \rho(\tilde{R}_C) \tilde{R}_C}{2\rho_p} \left(
 \frac{v_{\infty}^2 + 2GM_P(\tilde{R}_C) / \tilde{R}_C}{v_{\infty}^2 +
 2GM_P(\tilde{R}_C) / R_H}\right),
\end{equation}
where $v_{\infty}$ is the relative velocity of the planet and planetesimal when the two are far apart while $M_P(\tilde{R}_C)$ and $\rho(\tilde{R}_C)$ are the total mass of the planet and the density of the planet's envelope contained within $\tilde{R}_C$, respectively. Inaba \& Ikoma~(2003) propose replacing $\tilde{R}_C$ for $R_C$ in the expressions of collision probability, so $P_{coll}=P_{coll}(\tilde{R}_C,R_H, v_{rel})$. 

Planetesimal's relative velocities out of equilibrium were prescribed. The relative velocity, $v_{rel}$, between a planetesimal and the protoplanet may be described by
\begin{eqnarray}
v_{rel}= \sqrt{\frac{5}{8}e^2 + \frac{1}{2}i^2} v_k, 
\end{eqnarray} 
where $v_k$ is the keplerian velocity. We considered that the relative velocity is governed by two factors, the gravitational stirring caused by the protoplanets and the gas drag damping. We adopted the prescriptions developed by Ohtsuki et al.~(2002) for the gravitational stirring and the prescriptions developed by Adachi et al. (1976) for the gas drag damping (see Guilera et al., 2010, for a more detailed explanation).

Finally, the equations governing the evolution of the
gaseous envelope are those classic of stellar evolution theory 
\begin{eqnarray}
\frac{\partial r}{\partial m_r} & = & \frac{1}{4\pi r^2
 \rho}\qquad \mbox{equation of definition of mass}
\\
\frac{\partial  P}{\partial  m_r} &  = &  -  \frac{G
m_r}{4\pi  r^4}\qquad \mbox{equation of hydrostatic equilibrium}
\\ 
\frac{\partial L_r}{\partial m_r} & = & \epsilon_{pl}-T\frac{\partial
 S}{\partial t}\qquad \mbox{equation of energetic balance} 
 \\
\frac{\partial T}{\partial m_r} &  = & - \frac{G m_r T}{4\pi r^4 P}
\nabla\qquad \mbox{equation of energy transport,}
\end{eqnarray}
where $\rho$ is the density of the envelope, $G$ is the universal gravitational constant, $\epsilon_{pl}$ is the energy release rate due to the accretion of planetesimals, S is the entropy per unit mass, and $\nabla \equiv \frac{d \ln T}{d \ln P}$ is the dimensionless temperature gradient, which depends on the type of energy transport.

These equations were solved coupled self-consistently to the planetesimal's accretion rate (Eq.~\ref{eq:tas}), employing a standard finite difference (Henyey) method and a detailed constitutive physics as described in Fortier et al. (2007; 2009), and Guilera et al. (2010). 

In the following sections we shall present our results for the 
simultaneous, {\it in situ} formation of the four giant planets of the
solar system. Note that planets were assumed to be in circular orbits around the
Sun with fixed orbital radii consistent with the Nice model. Then, the migration of the planets is not addressed in the present study.

As we mentioned before, Thommes et al. (2008) showed that solar system analogues come out if gas giants do form but undergo modest migration and eccentricity growth, even for massive protoplanetary disks. Miguel et al. (2010 a; b) did a statistical study of the formation of planetary systems and they also found that solar system analogues are favored in massive disks where there is not a large accumulation of solids in the inner region of the disk, and only if type I migration is strongly reduced. Both works showed that toward low disk masses, planet formation is too slow to produce gas giants during the disk lifetime.

Several authors have argued that type I migration should be much slower than previously thought, otherwise planets get too close to or are even engulfed by the star  (see, for instance, Alibert et al. (2005) for detailed calculation of the formation of Jupiter and Saturn including migration). Tanaka et al. (2002) derived the migration rate for vertically isothermal disks. However, it was recently found that migration rates, and even the direction of migration, could be very different when vertically radiative or adiabatic transfer for the protoplanetary disk is adopted (Kley et al., 2009; Paardekooper et al., 2010). In light of recent results that suggest type I migration might not be as important as previously thought in more realistic disks (i.e not the idealized case proposed by Tanaka et al.), we find our hypothesis of {\it in situ} formation to be a good approximation to a more complex and realistic model, at least at early stages of the formation process. 

The neglect of type II migration is more problematic given that
this is likely to be an important effect in many systems, especially since the Nice model and its sequels calls for an early migration and multi-resonant capture of the giant planets. However, calculating type II migration rates is a complicated and subtle procedure in multi-planet systems and it is beyond the scope of this paper. It is important then to make here a note of caution: our results should be analyzed in the context of the {\it in situ} formation, which could be considered as a good approximation during the first stage of the formation of giant planets, i.e. before the gaseous runaway begins.

In addition to neglecting migration, our code does not calculate
the final stages of the gas accretion. Our code calculates the gas
accretion rate of the planet self-consistently using an adapted
Henyey-type code of stellar evolution, but always under the assumption
that the disk can supply the necessary amount of gas required by the
planet. Then, after the runaway growth of the envelope sets in, this
hypothesis may not be any more valid. However, recent works have shown
that gas accretion onto a critical mass core is likely to be
rapid. Machida et al. (2010) employing three-dimensional hydrodynamics
simulations, found that the gas accretion time-scale of a giant planet
is about $10^5$~yr (two orders of magnitude less than usually estimated the time-scale to
form the core). This means that the real bottleneck for giant planet
formation is the growth of a critical mass core in the first place. 

For the above mentioned reasons we think our code is suitable for studying the first stage of planetary formation: the growth of the planets up to their critical masses and how the simultaneous formation differs from the isolated case.


\section{Simultaneous formation of the solar system giant planets for a disk
with a
steep profile: $\Sigma \propto a^{-2}$}

\label{sec:sec3}

\begin{centering}
\begin{table}
\caption{Isolated formation of the solar system giant planets' cores for a disk
with surface densities of solids and gas $\propto a^{-2}$. Here $r_p^{min}$
stands for  the minimum radius of the size distribution of planetesimals. $t_{over}^{cross}$ represents the time at which the mass of the envelope of the planet  achieves the core's mass of the planet and gaseous runaway starts. $M_{over}^{cross}$ is the mass of the core at the cross-over time. Quantities between brackets, for the cases of Neptune and Uranus, correspond to the time, and the respective core mass, at which Neptune and Uranus achieve its current masses ($\sim 17$ and $\sim 14.5$ earth masses, respectively). When the cross-over time exceeds $\sim10$ Myr no values are given.}
\begin{tabular}{p{0.4cm}|p{0.58cm} p{0.58cm}|p{0.58cm} p{0.65cm}|p{0.58cm}
p{0.58cm}|p{0.58cm} p{0.65cm}}
\toprule[0.8mm]

 & Jupiter & & Saturn & & Neptune & & Uranus & \\
 $r_p^{min}$& $M_{over}^{cross}$ & $t_{over}^{cross}$ & $M_{over}^{cross}$& $t_{over}^{cross}$ & $M_{over}^{cross}$ & $t_{over}^{cross}$ & $M_{over}^{cross}$ & $t_{over}^{cross}$ \\
$[m]$ & $[\mathrm{M_{\oplus}}]$ & $[Myr]$ & $[\mathrm{M_{\oplus}}]$ & $[Myr]$ &
$[\mathrm{M_{\oplus}}]$ &
$[Myr]$ & $[\mathrm{M_{\oplus}}]$ & $[Myr]$ \\

 \midrule[0.6mm]

10 & --- & --- & 19.13 & 6.96 & 22.02 & 2.90 & 18.75 & 6.01 \\  

   &     &     &       &      & (14.62) & (2.31) & (12.59) & (4.98) \\  

\midrule[0.3mm]  

12.5 & --- & --- & 19.22 & 5.15 & 21.07 & 3.04 & 13.10  & 13.88  \\  

     &     &     &       &      & (14.57) & (2.43) & (11.24) & (12.39) \\  

\midrule[0.3mm]  

15 & --- & --- & 19.65 & 3.91 & 19.26 & 3.85 & --- & --- \\

   &     &     &       &      & (14.27) & (3.17) &   &  \\  

\midrule[0.3mm]  

20 & --- & --- & 20.90 & 2.83 & 14.48 & 8.69 & --- & --- \\

   &     &     &       &      & (12.78) & (7.82) &   &  \\  

\midrule[0.3mm]  

25 & 12.54 & 14.98 & 20.92 & 2.78 & --- & --- & --- & --- \\

\midrule[0.3mm]  

40 & 16.89 & 6.09 & 16.30 & 5.73 & --- & --- & --- & --- \\

\midrule[0.3mm] 

50 & 18.34 & 4.38 & 12.87 & 10.43 & --- & --- & --- & --- \\

\midrule[0.3mm] 

75 & 18.80 & 3.91 & --- & --- & --- & --- & --- & --- \\

\midrule[0.3mm] 

100 & 17.59 & 4.79 & --- & --- & --- & --- & --- & --- \\

\midrule[0.3mm] 

150 & 15.22 & 7.06 & --- & --- & --- & --- & --- & --- \\

\midrule[0.3mm] 

200 & 13.61 & 9.43 & --- & --- & --- & --- & --- & --- \\

\bottomrule[0.8mm] 
\end{tabular}
\label{table:tabla1} 
\end{table} 
\end{centering}

\begin{centering}
\begin{table}
\caption{Same as Table~\ref{table:tabla1} but for the simultaneous formation.}
\tiny{
\begin{tabular}{p{0.35cm}|p{0.61cm} p{0.625cm}|p{0.55cm} p{0.55cm}|p{0.61cm}
p{0.625cm}|p{0.61cm} p{0.625cm}}
\toprule[0.8mm]

 & Jupiter & & Saturn & & Neptune & & Uranus & \\
 $r_p^{min}$& $M_{over}^{cross}$ & $t_{over}^{cross}$ & $M_{over}^{cross}$& $t_{over}^{cross}$ & $M_{over}^{cross}$ & $t_{over}^{cross}$ & $M_{over}^{cross}$ & $t_{over}^{cross}$ \\
$[m]$ & $[\mathrm{M_{\oplus}}]$ & $[Myr]$ & $[\mathrm{M_{\oplus}}]$ & $[Myr]$ &
$[\mathrm{M_{\oplus}}]$ &
$[Myr]$ & $[\mathrm{M_{\oplus}}]$ & $[Myr]$ \\

 \midrule[0.6mm]

10 & $\sim 0.5$ & $\gg 10$ & 17.73 & 8.73 & 19.81 & 5.16 & 13.61 & 10.26 \\  

   &            &          &       &      & (14.86) & (4.28) & (11.80)  & (8.24)  \\ 

\midrule[0.3mm] 

50 & 15.47 & 8.89 & 13.42 & 10.44 & $\sim 1$ & $\gg 10$ & $\sim 0.5$ & $\gg
10$\\

\bottomrule[0.8mm] 
\end{tabular}
}
\label{table:tabla2} 
\end{table} 
\end{centering}

Adopting the initial configuration proposed by the Nice model, Desch (2007)
derived a surface density profile for an analogue of the 
minimum mass solar nebula which can be adjusted with a single power law, 
$\Sigma \propto a^{-2.168}$. This profile is much steeper than the one derived
by Hayashi (1981), where $\Sigma \propto a^{-1.5}$. We started then  our
investigation adopting a similar density profile as that proposed by Desch; in
our case, and for the sake of simplicity, we adopted 
$\Sigma \propto a^{-2}$. Pollack et al.~(1996) have used this profile to study the isolated formation of giant planets in solar system, but adopting a faster time-dependent regime for the growth of the core than the corresponding to the oligarchic growth regime.

Following the work of Benvenuto et al. (2009), we imposed the value of $11~\mathrm{g~cm^{-2}}$ for the initial solids surface density at the location of Jupiter. Then, at $t=0$ when calculations are started, the density profile throughout the disk is described by:
\begin{eqnarray}
\Sigma_s= 11 \, \left( \frac{a}{5.5~\mathrm{AU}}\right)^{-2} ~\eta_{ice}
\,\, \,\,\,\mathrm{g~cm^{-2}}  ,
\label{eq:ini_dens_sol}
\end{eqnarray}
where $\eta_{ice}$ takes into account the condensation of volatiles beyond the
snow line, considered to be located at 2.7~AU,
\begin{eqnarray}
\eta_{ice}= \left\lbrace
\begin{array}{ll}
1 & \quad a > 2.7~\mathrm{AU},\\
\frac{1}{4} & \quad a < 2.7~\mathrm{AU}.
\end{array}
\right.
\end{eqnarray}
In this work, we adopted a value for the gas-to-solid ratio of $Z_0^{-1} \simeq 65$, where $Z_0= 0.0153$ (Lodders et al., 2009) is the initial abundance of heavy elements in the Sun, implying that the initial surface density of gas is given by:
\begin{eqnarray}
\Sigma_g= 719 \left( \frac{a}{5.5~\mathrm{AU}}\right)^{-2}~\mathrm{g~cm^{-2}}.
\label{eq:ini_dens_gas}
\end{eqnarray}
Rescaling to 1~AU and spreading the snow line in a region of about 1 AU with a
smooth function as proposed by Thommes et al. (2003), the disk was defined
as: 
\begin{eqnarray}
\Sigma_s(a) &=&  \left\lbrace 83.19 + \left( 332.75-83.19\right)
\left[\frac{1}{2}\tanh\left(\frac{a-2.7}{0.5}\right) + 
\frac{1}{2}\right]\right\rbrace \times \nonumber \\
&&\left( \frac{a}{\mathrm{1 AU}}\right) ^{-2}~\mathrm{g~cm^{-2}},  \\
\Sigma_g(a) &=& 21750 \left(\frac{a}{\mathrm{1~AU}}
\right)^{-2}~{\mathrm{g~cm^{-2}}}.
\end{eqnarray}
Regarding the temperature profile, we adopted the same one as in our previous
works,
\begin{eqnarray}
T(a) &=& 280 \left(\frac{a}{\mathrm{1~AU}}\right)^{-1/2}~\mathrm{K}.
\end{eqnarray}
We did not consider the evolution in time of the temperature profile, which implies
that the temperature of the nebula at a given location is a fixed external
boundary condition.
The initial mass of the disk was $\sim 0.066~\mathrm{M_{\odot}}$, this value
being the
result of integrating the surface density profile from $a=0.4$~AU to
$a=30$~AU. 

With the disk so defined, we first calculated the isolated formation of each
 of the four giant planets of the solar system.
For each planet we run several simulations only changing  the minimum radius
 of the size distribution of planetesimals, $r_p^{min}$. The aim of this procedure
was to look for an interval in the planetesimals radii
where the isolated formation of all the planets occurs in less than 10~Myr.
Afterwards, and using these results as a guide, we looked for an optimum value of
$r_p^{min}$ to calculate the
simultaneous formation of the four planets.

\begin{figure} 
\centering 
\includegraphics[width= 0.35\textwidth, angle= 270]{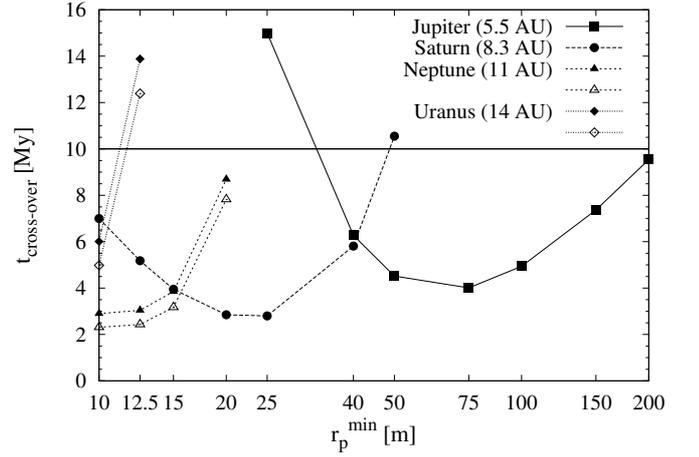} 
\caption{Cross-over time as function of the minimum radius of the planetesimals
size distribution for the case of the isolated formation of each planet. Results
correspond to a disk with a surface density profile $\Sigma \propto a^{-2}$. Open triangles (diamonds) correspond to the time at which Neptune (Uranus) achieves its current mass ($\sim 17$ and $\sim 14.5$ Earth masses, respectively).}
\label{fig:fig1} 
\end{figure}

\begin{figure} 
\centering 
\includegraphics[width= 0.35\textwidth, angle= 270]{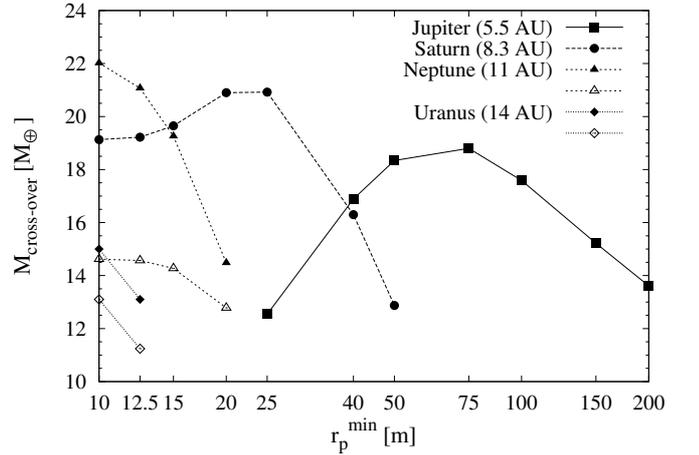} 
\caption{Core mass at the cross-over time as function of the minimum radius of the planetesimals size distribution for the case of the isolated formation of each planet. Results correspond to a disk with a surface density profile $\Sigma \propto a^{-2}$. Open triangles (diamonds) correspond to mass of the core at the time in which Neptune (Uranus) achieves its current mass ($\sim 17$ and $\sim 14.5$ Earth masses, respectively).}
\label{fig:fig2} 
\end{figure}

The results for the isolated formation are summarized in Table~\ref{table:tabla1}. We found that each planet has an independent interval for $r_p^{min}$ in which the cross-over time (the time at which the mass of the envelope achieves the core's mass, and gaseous runaway starts) is reached in less than
10~Myr (Fig.~\ref{fig:fig1}). Jupiter reached its cross-over mass, and could be formed, in less than $\sim 10$~Myr for $r_p^{min}$ in the range of $\sim 30$ and 200~m. Saturn, on the other hand,
reached its cross-over mass in less than $\sim 10$~Myr for  $r_p^{min}$ between 10 and $\sim 50$~m. In the case of Neptune, $r_p^{min}$ had to be between 10 and 20~m. Finally, Uranus could only be formed in agreement with the time restriction for $r_p^{min} \lesssim 12.5$~m. 

Note that we did not consider radii smaller than 10~m. This is due to the fact that in most cases (see next Sections) for such small planetesimals the time-scale of formation of the outermost planets are significantly lower than the time-scale corresponding for Jupiter and Saturn. We consider very unlikely that Uranus and Neptune reached their current masses before Jupiter and Saturn reached their cross-over masses, as in that case they could have continue their growth and become gas giants.

For Desch's steep profile, we found that the shortest cross-over time for Jupiter
is $\sim 4$~Myr, for Saturn is $\sim 2.8$~Myr, for Neptune is $\sim 2.9$~Myr
and for Uranus is $\sim 6$~Myr. However, these time-scales did not occur for the same
minimum radii of the planetesimals populating the disk.

On the other hand, we found that for Jupiter and Saturn the cross-over masses were in good agreement with the current theoretical estimations of their solid content (Fig.~\ref{fig:fig2}). Note that we assumed that all the infalling planetesimals reach the core's surface without losing mass on their trajectories throughout the envelope, this
meaning that $M_c$ corresponds in reality to the total heavy element's mass in the interior of the planet (core {\it plus} solids in the envelope). Theoretical
models estimate the core mass of Jupiter between $0 -
12~\mathrm{M_{\oplus}}$
(Guillot, 2005) or $14 - 18~\mathrm{M_{\oplus}}$ (Militzer et al., 2008),
depending on the equation of state employed. Regarding the
core mass of Saturn, it is accepted to be in the range of  $9 -
22~\mathrm{M_{\oplus}}$ (Guillot, 2005). 
Models also predict $10 - 40~\mathrm{M_{\oplus}}$ and $20 -
30~\mathrm{M_{\oplus}}$ of
total heavy elements (envelope $+$ core) in Jupiter and Saturn, respectively
(Guillot, 2005;
Guillot \& Gautier, 2009). On the other hand, Uranus and Neptune are mostly
rocks
and ices. The planetary interior models of Podolak et al., (2000) place upper
limits on the
H/He contents of Uranus and Neptune, being for Uranus of 4.2 and
3.2~$\mathrm{M_{\oplus}}$ for Neptune.
If only hydrogen and helium are present in the atmosphere, a lower limit for the
gas mass of
0.5~$\mathrm{M_{\oplus}}$ is obtained for each planet (Guillot, 2005). 
To relate the above mentioned values to our results of the core mass, in the
following we will
consider that the total amount of heavy elements contained in each planet
corresponds to the mean value of the current estimated boundaries. This means
that we are going to compare our results for the mass of the core
with: $25\pm15~\mathrm{M_{\oplus}}$ for Jupiter, $25\pm5~\mathrm{M_{\oplus}}$
for Saturn, 
$15.15\pm1.35~\mathrm{M_{\oplus}}$ for Neptune, and
$12.15\pm1.85~\mathrm{M_{\oplus}}$, for Uranus.

Regarding the formation of Neptune and Uranus, we can see (Table~\ref{table:tabla1} and Fig.~\ref{fig:fig2}) that if we let them grow without restriction concerning their masses, for most cases their cross-over masses were higher than their current masses. However, the masses of the cores at the time they reached their current masses are in agreement with theoretical estimations.

For the cases of Jupiter and Saturn, we found that there was an optimum value
for $r_p^{min}$ of around 70~m for Jupiter, and between $20 - 25$~m for Saturn, for which the accretion was most effective. The existence of an optimum $r_p^{min}$ for
the size distribution of the planetesimal radii is due to the fact that
planetesimal's migration velocities are $\propto 1/r_p$ (Eq. \ref{eq:vel-migra}) and that the
planetesimal's accretion by the protoplanet is more efficient for small objects. Then, there is a competition between the
efficiency
of the accretion and the efficiency of the planetesimal migration. As we
considered a planetesimal mass distribution that follows the power
law $n(m)= m^{-2.5}$ (where most of the mass of solids lie in the small objects),
$r_p^{min}$ has to be chosen considering the above mentioned compromise between the migration and the accretion rate. If the radii of the smaller planetesimals are very small, most of the mass of solids would pass through the feeding zone of the planet without being accreted. If, on the other hand, they are very large, the accretion time scale would be longer than the protoplanetary disk lifetime.

The reason why the intervals for $r_p^{min}$ of each planet are different is
due to the fact that  both the accretion rate of planetesimals and their
migration velocities vary with the distance to the central star. For this steep
profile,
we did not find an overlap between these intervals, so it seems that there is
not
a common size distribution for the planetesimal radii that would allow us to
form
simultaneously the four planets in less than 10~Myr. 

However, in spite of the previous results, we checked if effectively there was
not an
interval where the simultaneous formation could occur. Taking into account the
results for the calculations of isolated formation, we run some simulations
where we considered the simultaneous formation of the four giant planets.

\begin{figure} 
\centering 
\includegraphics[width= 0.35\textwidth, angle= 270]{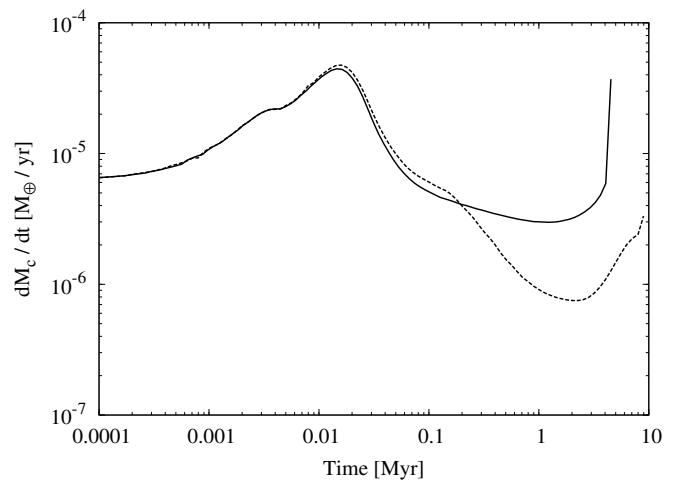}
\caption{Time evolution of Jupiter's planetesimals accretion rate for the cases of isolated (solid line) and simultaneous formation (dashed line) for a disk profile $\Sigma \propto a^{-2}$.}
\label{fig:fig2bis} 
\end{figure}

\begin{figure}
\centering 
\includegraphics[width= 0.35\textwidth, angle= 270]{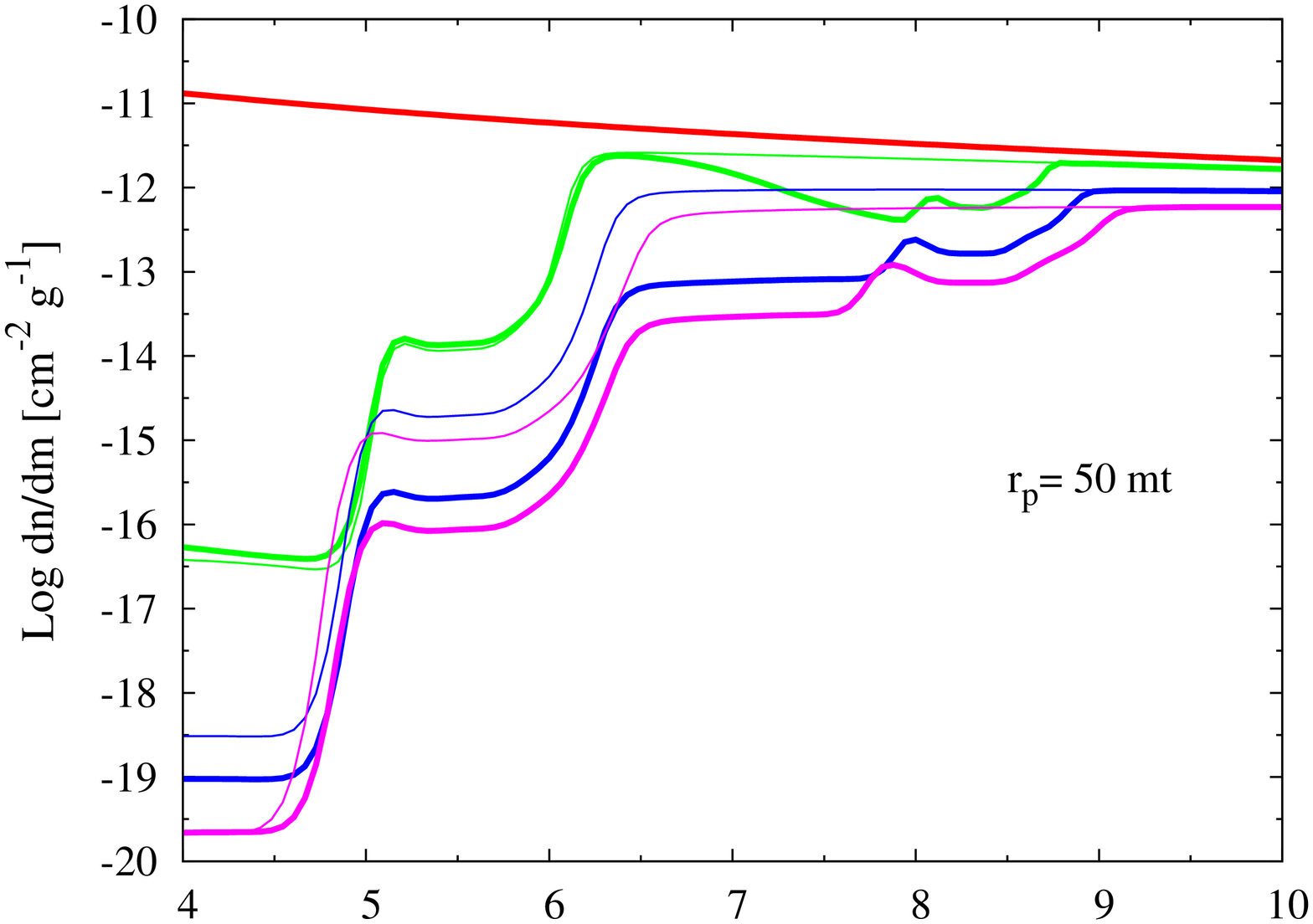}
\includegraphics[width= 0.35\textwidth, angle= 270]{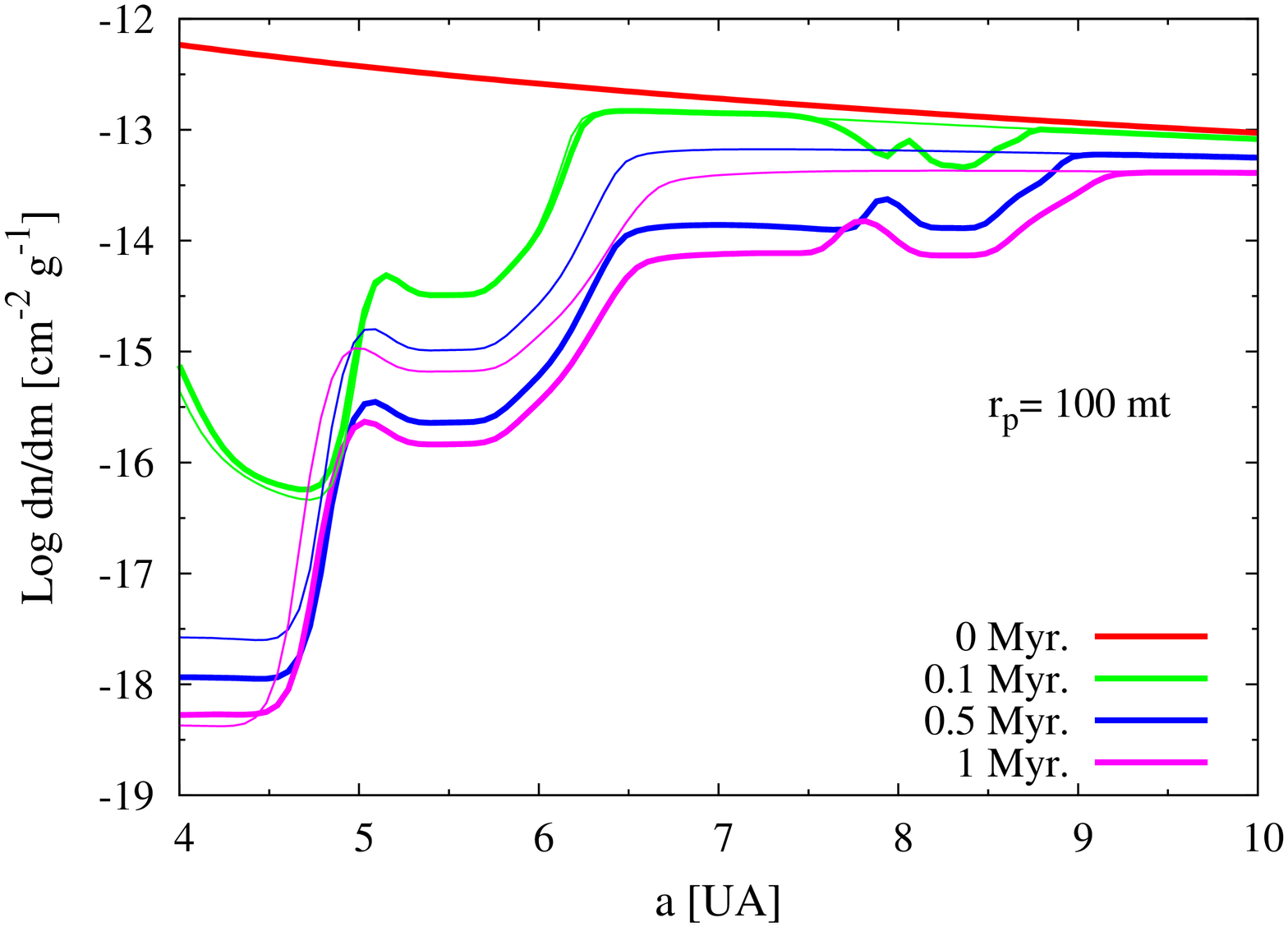}
\caption{Time evolution of the surface density of solids for small planetesimals in a disk with power index $p= 2$. Thin lines correspond to the isolated formation of Jupiter, thick lines correspond to the simultaneous formation of the giant planets. We can see how the presence of Saturn significantly reduces the surface density of solids at Jupiter's location by the accretion of small planetesimals. The presence of Saturn delayed the formation time of Jupiter by a factor $\sim 2$ (color figure only available in the electronic version).}
\label{fig:fig2-3} 
\end{figure}

Results are listed in Table~\ref{table:tabla2}. We first chose $r_p^{min}= 50$~m for the size distribution of planetesimals. In this case the formation times for Jupiter and
Saturn were of $\sim 9$~Myr and $\sim 10.5$~Myr, respectively. However, in this
time interval the embryos of Neptune and Uranus achieved only a core mass of  $M_c \sim
1~\mathrm{M_{\oplus}}$ and $M_c \sim 0.5~\mathrm{M_{\oplus}}$ respectively, both with a
negligible envelope (the simulation was stopped at $\sim 11$~Myr, when Saturn finished its formation). We note that the presence of Saturn delayed the formation time of Jupiter by a factor $\sim 2$ when we compare this results to the isolated ones. As we can see in Fig.~\ref{fig:fig2bis}, the presence of Saturn significantly decreased the solid accretion rate of Jupiter. The reduction in the solid accretion rate is due to the fact that planetesimals that were accreted by Jupiter in the isolated case (coming from the outer region of the Solar System), were accreted first by Saturn when the formation of both planets was calculated simultaneously. For this steep profile, the presence of Saturn acts as a sink of planetesimals, significantly decreasing the surface density of smaller planetesimals in the neighborhood of Jupiter when we compare it with the isolated formation of Jupiter (see Fig. \ref{fig:fig2-3}).  

When we considered smaller planetesimals ($r_p^{min}= 10$~m), Neptune turned out to be the first one to achieve its cross-over mass, followed by Saturn and Uranus, while Jupiter was not able to reach its cross-over mass in less than 10 Myr (the simulation was stopped at $\sim 15$~Myr and Jupiter's core mass was only $\sim 0.5~\mathrm{M_{\oplus}}$ with almost no gas bound). However, in the simultaneous case, the formation of all the planets took longer than in the isolated case. We consider that a case like this is very unlikely because, while Jupiter could not grow up to its cross-over mass, Neptune reached its cross-over mass before Saturn, which means that it could have  started the gaseous runaway (and become a gas giant) before Saturn completed its formation. While it is widely accepted that Uranus and Neptune did not achieve the gaseous runaway to become gas giants, we remark that in our simulations we did not halt the accretion onto them when they reach their current masses and we allowed Neptune and Uranus to grow up to the cross-over mass, as the perturbations that a planet produces on the planetesimal disk, and how it affects the formation of the others, depends on its mass. This is in line with our working hypothesis of studying the formation of the planets up to their cross-over masses, a process that is well described with our code. In this work we do not attempt to explain the final masses of the planets.

Then, for a nebula like the one considered here we would
need that $r_p^{min}$ 
decreases with the distance from the central star in order to be able to form
the four planets simultaneously. Moreover, an accurate treatment demands that
the simultaneous formation should be recalculated with a non homogeneous
distribution.
 However, in principle, there's no physical reason to
support such a decrease in the minimum radii of the planetesimals as we move 
away from the Sun. In fact, a decrease in the planetesimals' radii would be in contradiction with the results found by Chambers~(2010), where planetesimal size should increase with distance.


\section{Simultaneous formation of the solar system giant planets for shallow
disk density profiles}

\label{sec:sec4}

In the previous Section we showed that with a steep profile, as the one derived
by Desch (2007), for a minimum mass solar nebula compatible with the Nice
model, there is not a minimum radius for the population of the planetesimals that favors the simultaneous 
formation of the four giant planets of the solar system. In this section we will
analyze other density profiles, all of them power laws where we change the value
of the index $p$.

\subsection{The case of $\Sigma \propto a^{-3/2}$}
\label{sec:sec4.1}

Employing the standard surface density profile $\Sigma \propto a^{-3/2}$ we
repeated the calculations of Sec.~\ref{sec:sec3}.
The same procedure as in the previous section was employed: we calculated first the
isolated formation in order to look for candidate values for $r_p^{min}$ and,
afterwards, with this values we calculated the simultaneous formation of the
four planets.

Imposing again the value of $11~\mathrm{g~cm^2}$ for the solid surface density
at the location of Jupiter, using the same solid/gas ratio, and 
rescaling to 1~AU, the disk is defined by
\begin{eqnarray}
\Sigma_s(a) &=&  \left\lbrace 35.5 + \left( 142-35.5\right)
\left[\frac{1}{2}\tanh\left(\frac{a-2.7}{0.5}\right) + 
\frac{1}{2}\right]\right\rbrace \times \nonumber \\
&&\left( \frac{a}{1~\mathrm{AU}}\right) ^{-3/2}~\mathrm{g~cm^{-2}},  \\
\Sigma_g(a) &=& 9281 \left(\frac{a}{1~\mathrm{AU}} \right)^{-3/2} \,\,
\mathrm{g~cm^{-2}}.
\end{eqnarray}  

According to this definition, the initial mass of the disk was $\sim
0.063~\mathrm{M_{\odot}}$.

\begin{centering}
\begin{table}
\caption{Same as Table~\ref{table:tabla1} but for a disk
with surface density of solids and gas $\propto a^{-3/2}$.}
\begin{tabular}{p{0.4cm}|p{0.58cm} p{0.58cm}|p{0.58cm} p{0.65cm}|p{0.58cm}
p{0.58cm}|p{0.58cm} p{0.65cm}}
\toprule[0.8mm]

 & Jupiter & & Saturn & & Neptune & & Uranus & \\  
$r_p^{min}$& $M_{over}^{cross}$ & $t_{over}^{cross}$ & $M_{over}^{cross}$& $t_{over}^{cross}$ & $M_{over}^{cross}$ & $t_{over}^{cross}$ & $M_{over}^{cross}$ & $t_{over}^{cross}$ \\
$[m]$ & $[\mathrm{M_{\oplus}}]$ & $[Myr]$ & $[\mathrm{M_{\oplus}}]$ & $[Myr]$ &
$[\mathrm{M_{\oplus}}]$ & $[Myr]$ & $[\mathrm{M_{\oplus}}]$ & $[Myr]$ \\

\midrule[0.6mm]

10 & --- & --- & 23.49 & 4.14 & 28.28 & 1.10 & 32.20 & 0.61 \\  

   &     &     &    &    & (15.30) & (0.75) & (13.67) & (0.35) \\

\midrule[0.3mm]  

20 & 19.88 & 9.85 & 26.93 & 1.33 & 30.33 & 0.87 & 26.15 & 1.91 \\

   &       &      &       &      & (15.58) & (0.61) & (13.37) & (1.45) \\

\midrule[0.3mm]

35 & 20.92 & 4.58 & 28.38 & 1.22 & 24.02 & 2.52 & 17.57 & 7.62 \\

   &       &      &       &      & (15.05) & (2.08) & (12.45) & (6.50) \\

\midrule[0.3mm]

40 & 21.65 & 3.76 & 27.53 & 1.44 & 22.33 & 3.28 & 15.19 & 10.92 \\

   &       &      &       &      & (14.85) & (2.77) & (11.90) & (9.56) \\

\midrule[0.3mm]

80 & 23.67 & 2.61 & 21.08 & 3.90 & 13.11 & 13.20 & --- & --- \\

   &       &      &       &      & (12.3) & (12.6) & --- & --- \\

\midrule[0.3mm] 

160 & 21.03 & 3.70 & 14.59 & 10.73 & --- & --- & --- & --- \\

\midrule[0.3mm] 

400 & 15.98 & 7.81 & --- & --- & --- & --- & --- & --- \\

\bottomrule[0.8mm] 
\end{tabular}
\label{table:tabla3} 
\end{table} 
\end{centering}

\begin{centering}
\begin{table}
\caption{Same as Table~\ref{table:tabla3} but for the simultaneous formation of
the four planets.}
\begin{tabular}{p{0.4cm}|p{0.58cm} p{0.58cm}|p{0.58cm} p{0.65cm}|p{0.58cm}
p{0.58cm}|p{0.58cm} p{0.65cm}}
\toprule[0.8mm]

 & Jupiter & & Saturn & & Neptune & & Uranus & \\
$r_p^{min}$& $M_{over}^{cross}$ & $t_{over}^{cross}$ & $M_{over}^{cross}$& $t_{over}^{cross}$ & $M_{over}^{cross}$ & $t_{over}^{cross}$ & $M_{over}^{cross}$ & $t_{over}^{cross}$  \\
$[m]$ & $[\mathrm{M_{\oplus}}]$ & $[Myr]$ & $[\mathrm{M_{\oplus}}]$ & $[Myr]$ &
$[\mathrm{M_{\oplus}}]$ &
$[Myr]$ & $[\mathrm{M_{\oplus}}]$ & $[Myr]$ \\

 \midrule[0.6mm]

20 & 35.74 & 0.99 & 38.30 & 0.55 &  33.65  &  0.60  & 27.83~* & 1.6~* \\
 
   &       &      &       &      & (16.20) & (0.42) & (13.92)  & (1.1)      \\

\midrule[0.3mm] 

35 & 31.89 & 2.25 & 29.12 & 1.83 & 23.43 & 3.88 & 16.74 & 8.68 \\

 &  &  &  &  & (15.52) & (3.01) & (12.69) & (6.80) \\

\midrule[0.3mm] 

40 & 30.08 & 2.96 & 27.28 & 2.55 &  20.72  &  5.27  &  14.70  &  12.23  \\

   &       &      &       &      & (15.12) & (4.19) & (12.18) & (10.02) \\

\bottomrule[0.8mm] 
\end{tabular}
\label{table:tabla4}
\begin{list}{}{}
\item[*]{For this case, the numerical models corresponding to Uranus did
	not converge (Uranus did not achieve its cross-over mass) and
	the simulation was halted. At time $t=1.6$~Myr., Uranus reached
	a core mass of $27.83~\mathrm{M_{\oplus}}$ with a corresponding
	envelope mass of $8.75~\mathrm{M_{\oplus}}$.} 
\end{list} 
\end{table} 
\end{centering}

The results obtained for the isolated formation of each planet are summarized in
Table~\ref{table:tabla3}. We note that these results are qualitatively similar
to those obtained for the steeper profile $\Sigma \propto a^{-2}$. However, in
this case, there
are some values for $r_p^{min}$ (between $20 - 40$~m) that may allow the 
simultaneous formation of all the planets with a common size
distribution of planetesimals. 

\begin{figure} 
\centering 
\includegraphics[width= 0.35\textwidth, angle= 270]{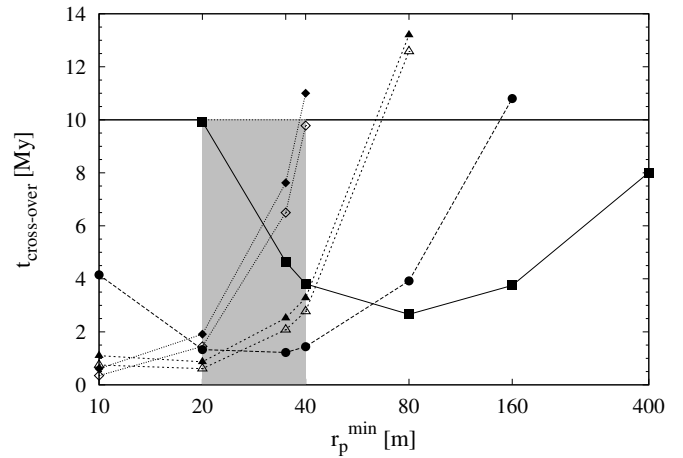} 
\caption{Same as Fig.~\ref{fig:fig1} but considering a disk with a surface
density profile $\Sigma \propto a^{-3/2}$. The gray zone corresponds to the interval of values of the minimum radius of the size distribution of planetesimals where the cross-over times corresponding to the four planets are less than $\sim 10$~Myr. Squares, circles, triangles and diamonds correspond to Jupiter, Saturn, Neptune and Uranus, respectively.}
\label{fig:fig3} 
\end{figure}

For the case of the isolated formation of each planet, Jupiter reached its cross-over mass in less than 10~Myr for $20~\mathrm{m} \lesssim r_p^{min} \lesssim 400~\mathrm{m}$, Saturn for $10~\mathrm{m} \lesssim r_p^{min} \lesssim
160~\mathrm{m}$, Neptune for $10~\mathrm{m} \lesssim r_p^{min} \lesssim
80~\mathrm{m}$, and Uranus for $10~\mathrm{m} \lesssim r_p^{min} \lesssim
40~\mathrm{m}$. Then, for this protoplanetary nebula, there should be an optimum radius $r_p^{min}$ for which the four planets achieved their cross-over masses in less than 10 Myr (Fig.~\ref{fig:fig3}), this radius lying in the interval [20, 40] m. Note that for a planetesimal's size distribution between $r_p^{min}= 20$~m and $r_p^{max}=100$~km, the first planet to reach its cross-over mass was Neptune, followed by Saturn, Uranus and Jupiter (see Table~\ref{table:tabla3}). However, for a size distribution of planetesimals between $r_p^{min}= 35$~m and $r_p^{max}$=100~km, and between $r_p^{min}= 40$~m and $r_p^{max}= 100$~km, the first planet to reach the cross-over mass was Saturn, followed by Neptune, Jupiter and Uranus.

\begin{figure}
\centering 
\includegraphics[width= 0.35\textwidth, angle= 270]{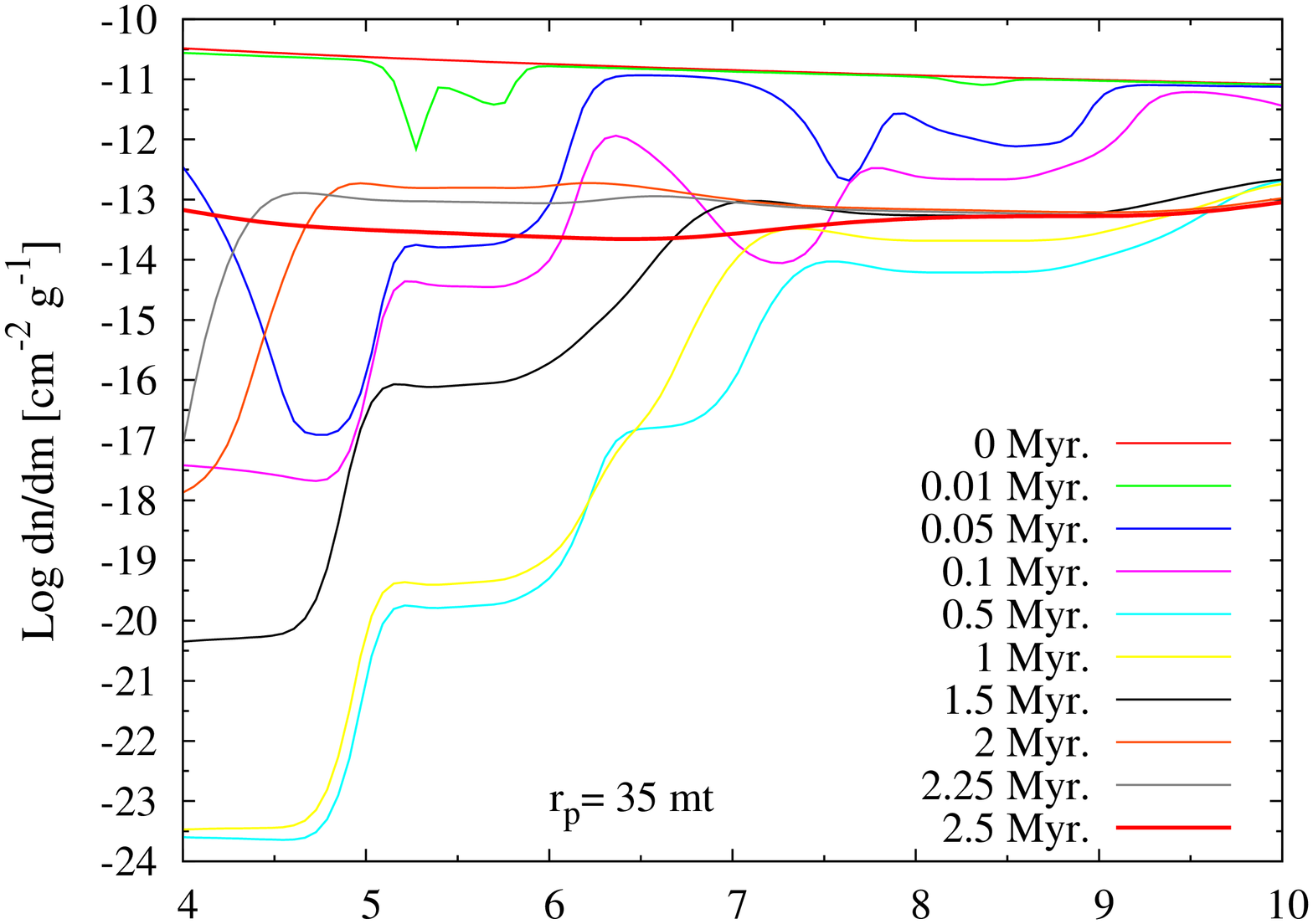}
\includegraphics[width= 0.35\textwidth, angle= 270]{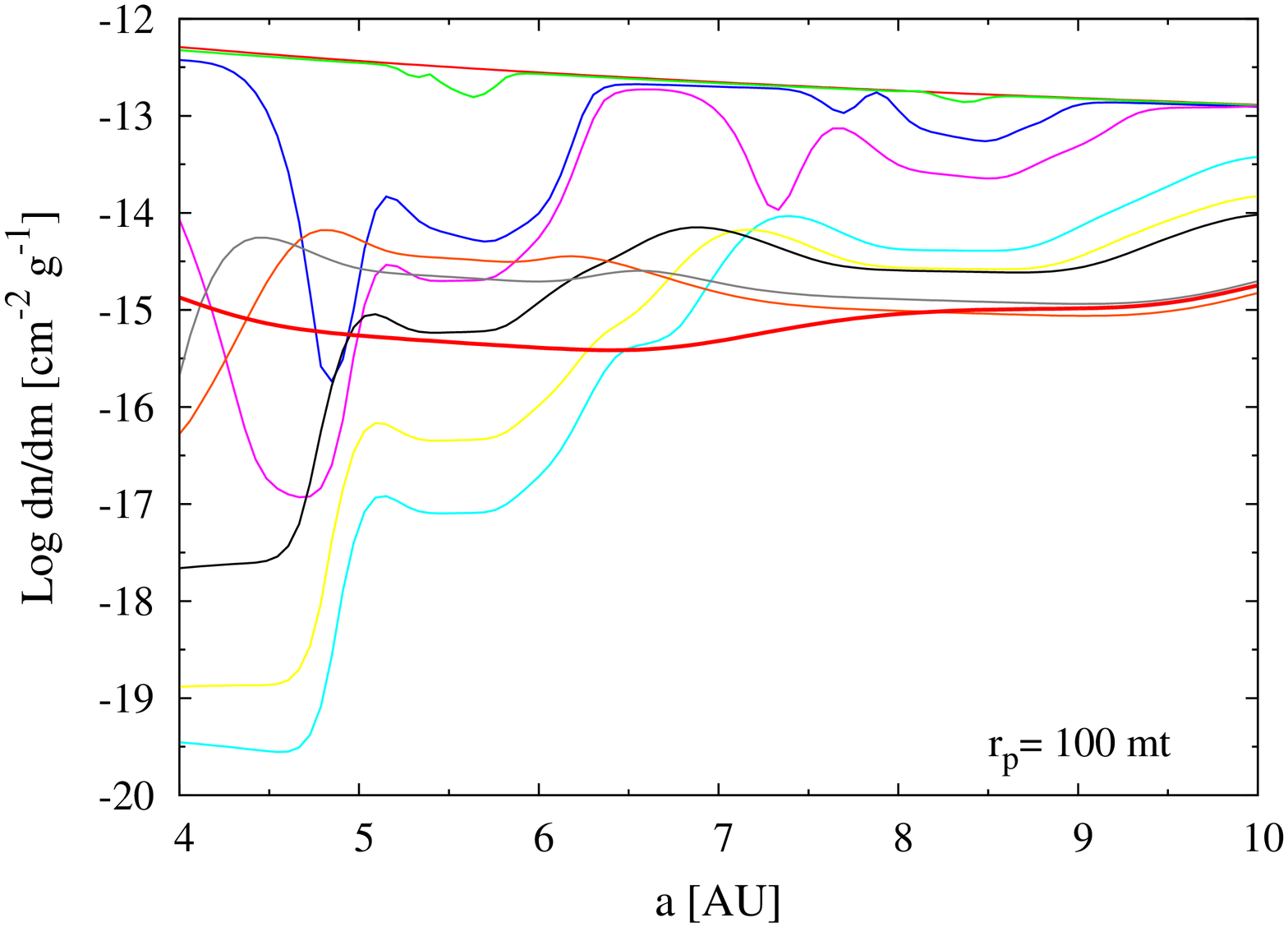}
\caption{Time evolution of the surface density of solids for small planetesimals at Jupiter-Saturn region, for a disk with a power index $p= 3/2$, considering the case of the simultaneous formation and employing a distribution of planetesimals between $r_p^{min}= 35$ and $r_p^{max}= 100$~km. After 0.5~Myr. the surface density of solids at Jupiter's neighborhood is significantly augmented by the incoming flow of planetesimals due to the gravitational perturbations of the outer planets (color figure only available in the electronic version).}
\label{fig:fig5-2} 
\end{figure}

\begin{figure}
\centering 
\includegraphics[width= 0.35\textwidth, angle= 270]{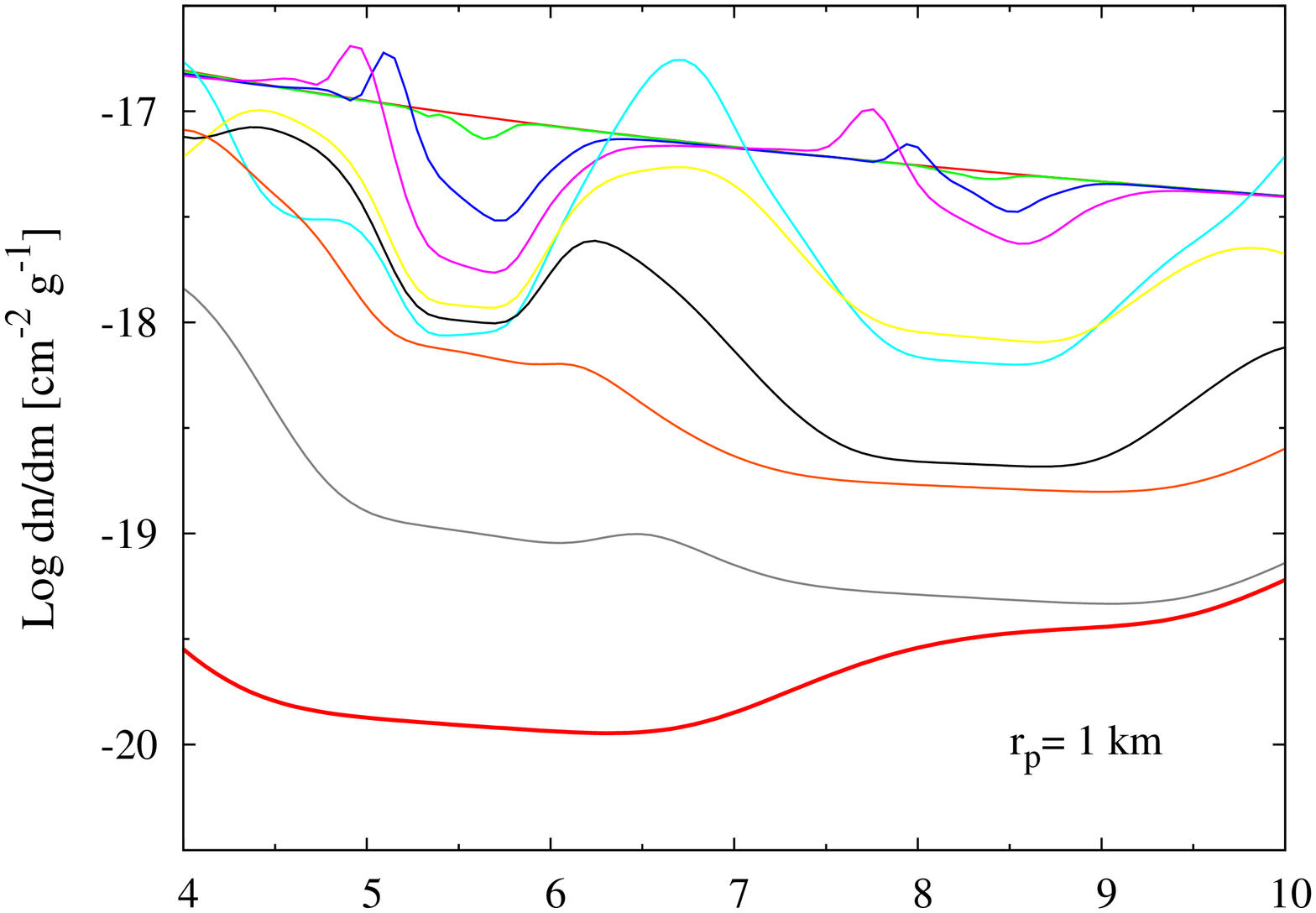}
\includegraphics[width= 0.35\textwidth, angle= 270]{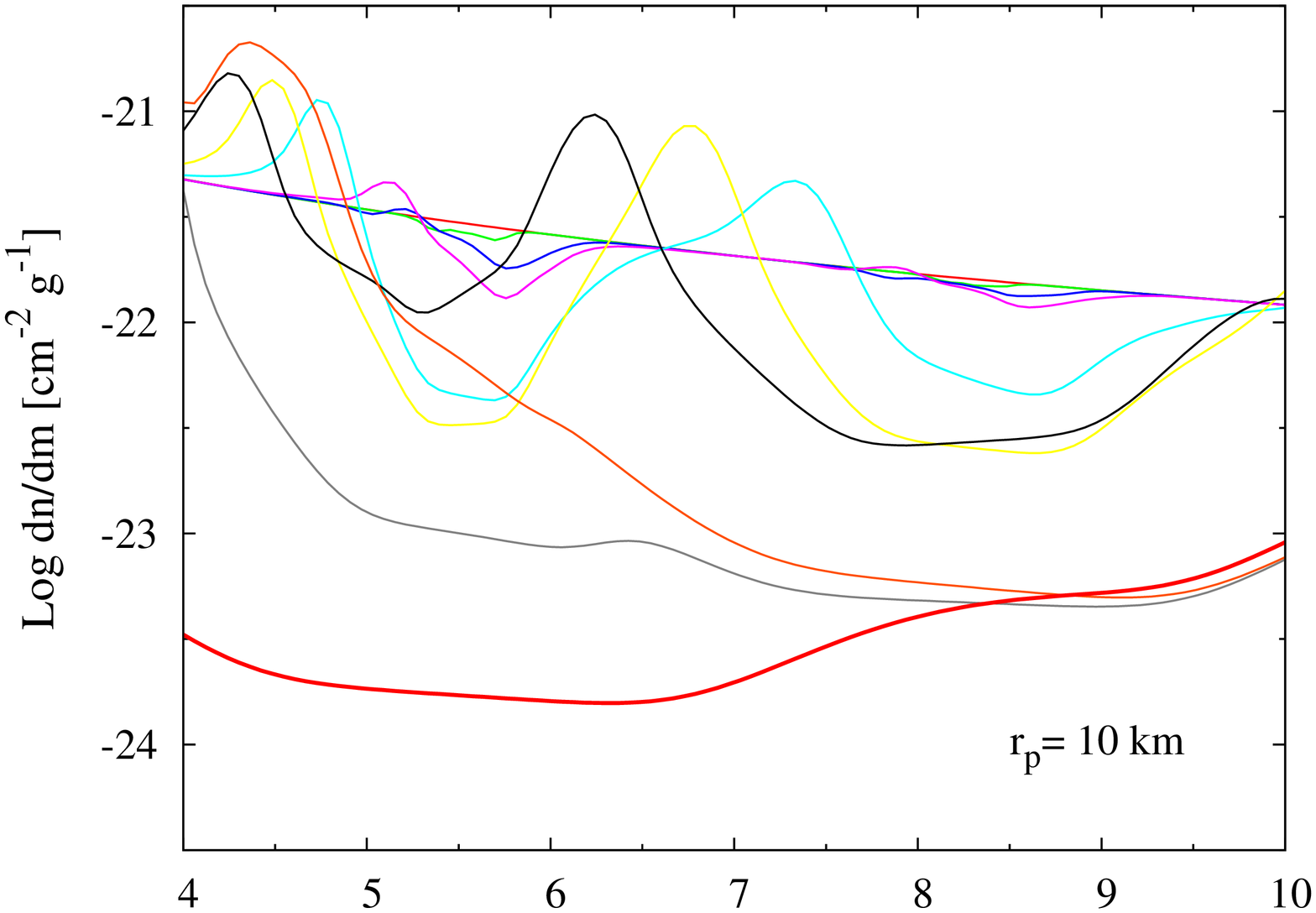}
\includegraphics[width= 0.35\textwidth, angle= 270]{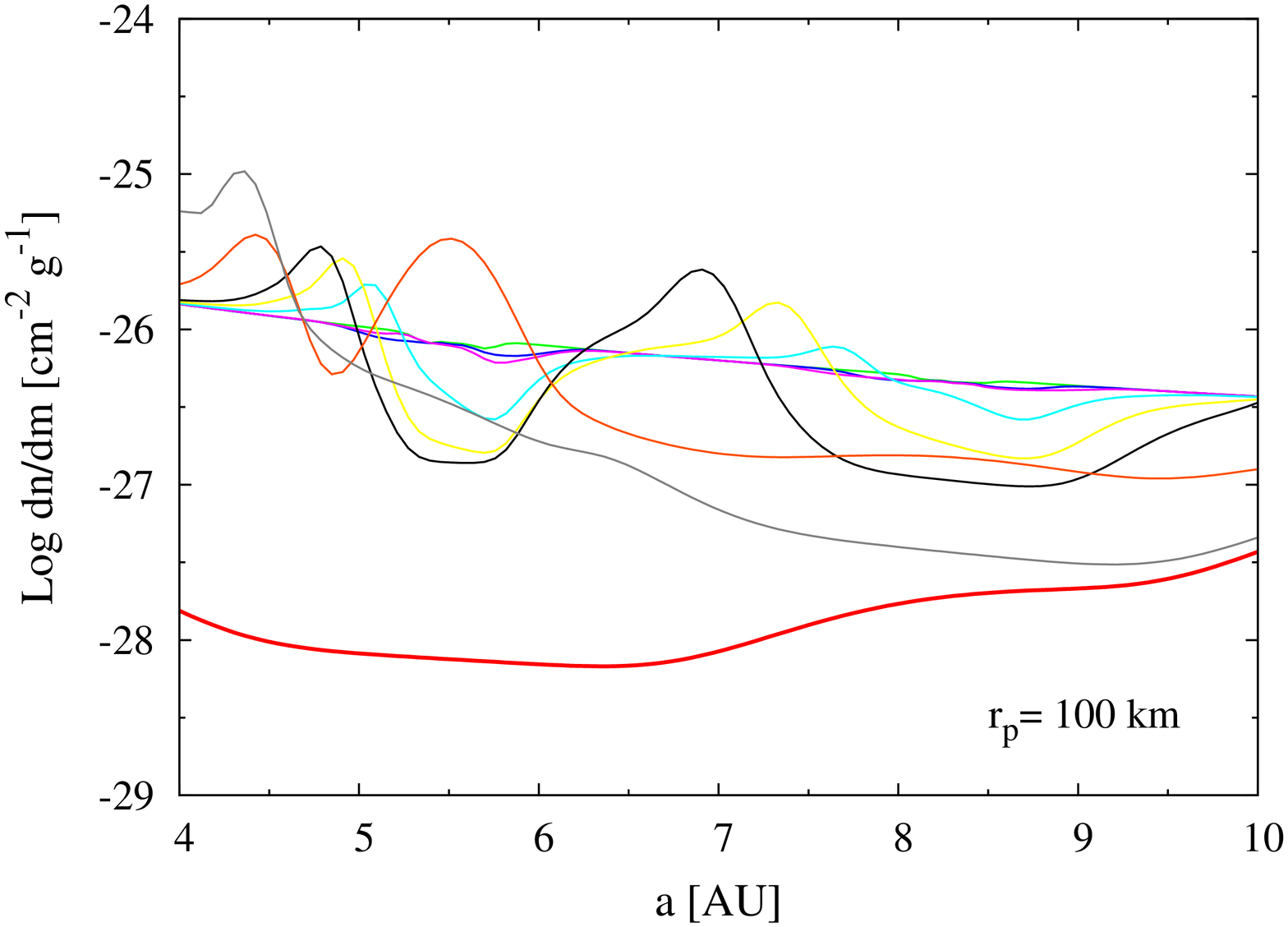}
\caption{Same as Fig.~\ref{fig:fig5-2} but for planetesimals with radii larger than 1~km. In this case the shape of the density wave is well defined. The density wave increases the surface density of solids to values that are even larger than the initial ones (color figure only available in the electronic version).}
\label{fig:fig5-3} 
\end{figure}

\begin{figure} 
\centering 
\includegraphics[width= 0.35\textwidth, angle= 270]{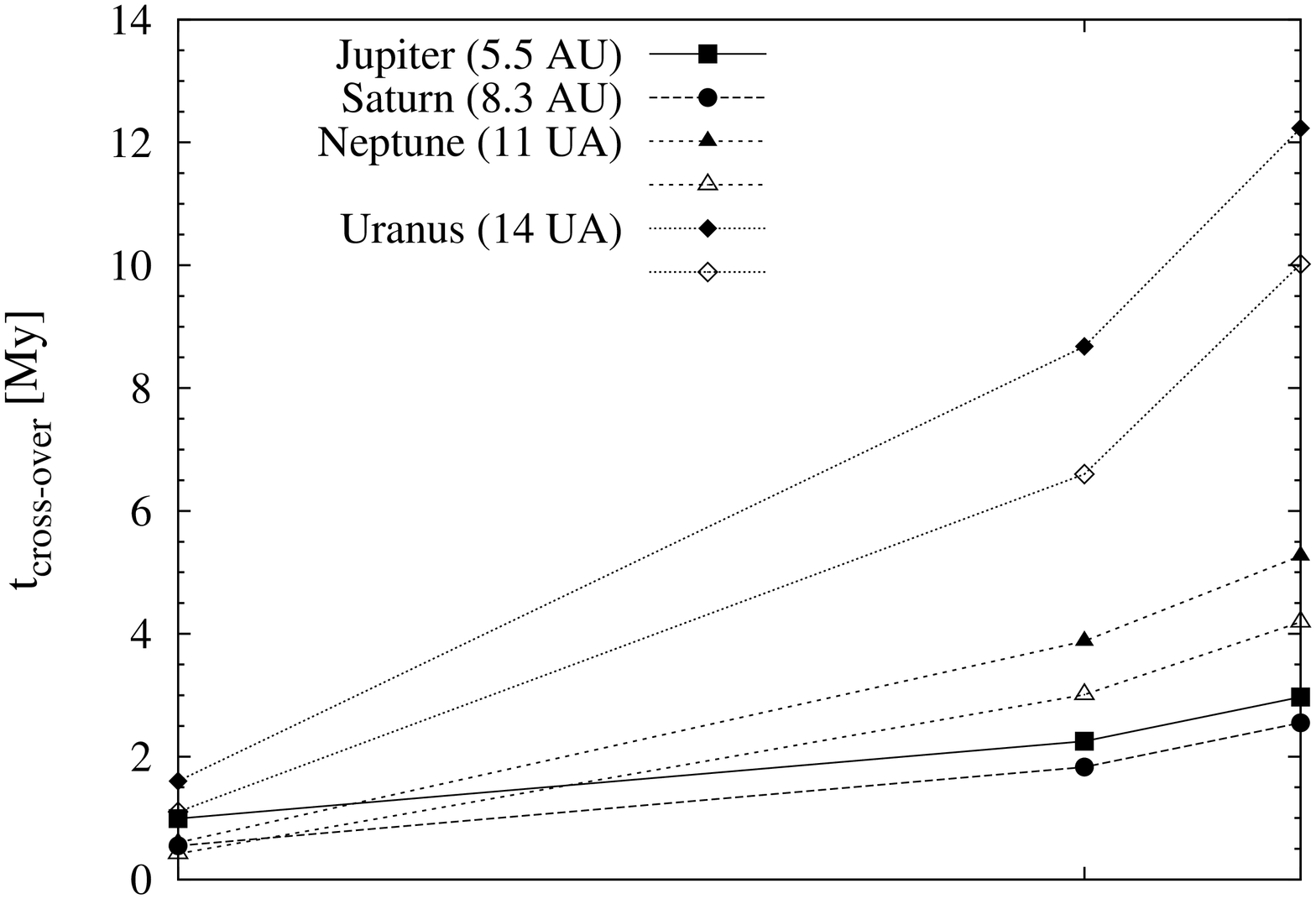}
\includegraphics[width= 0.35\textwidth, angle= 270]{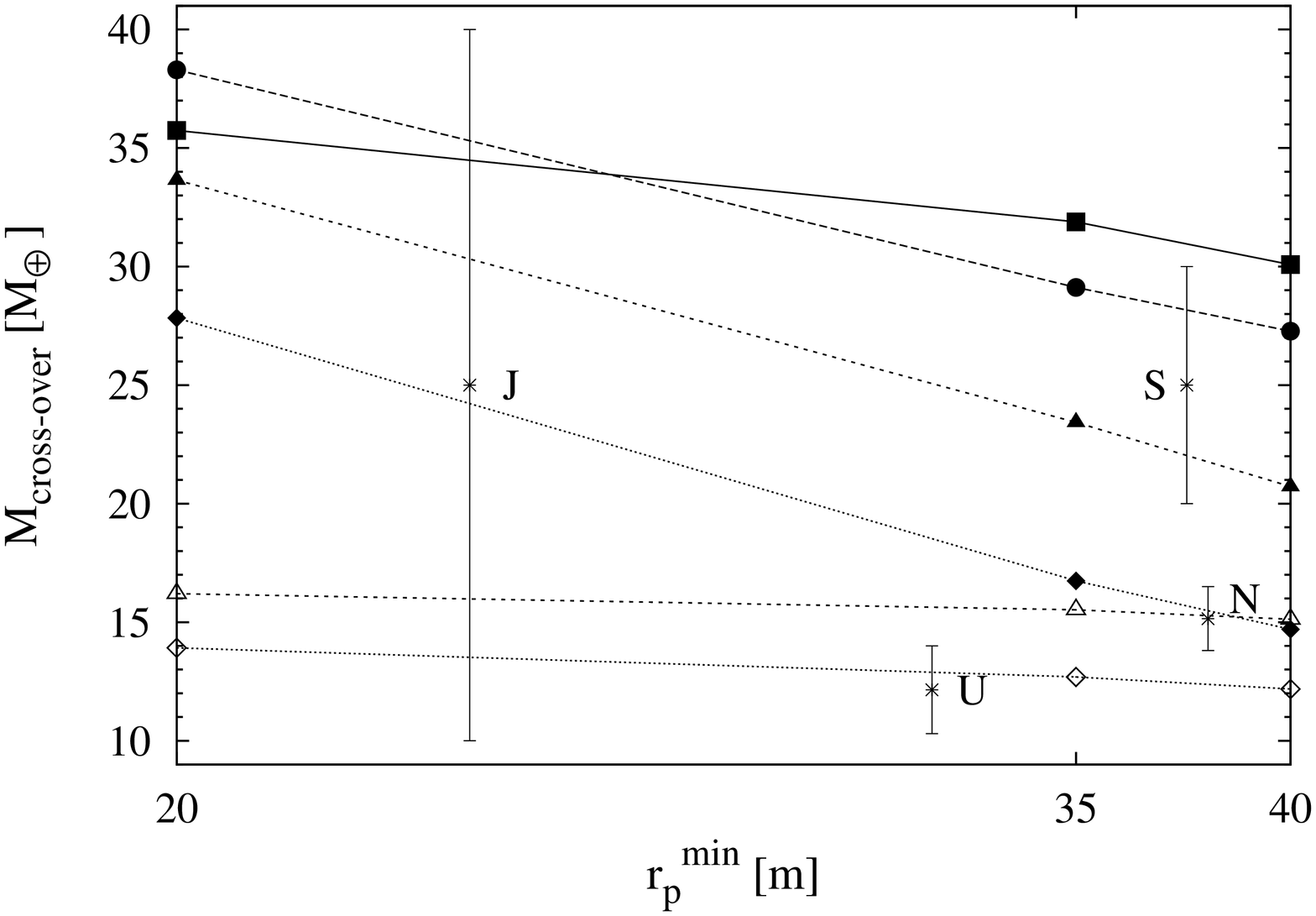}
\caption{Cross-over time (top) and cross-over mass (bottom) as function of the minimum radius of the size distribution of planetesimals for the case of the simultaneous formation. Here the disk surface density is characterized by $\Sigma \propto a^{-3/2}$. For $r_p^{min}= 35$ and 40~m the cross-over times and cross-over masses of Jupiter and Saturn are in good agreement with the observational and theoretical estimations. Regarding Neptune and Uranus their cross-over masses are higher than their current total masses. However, for both planets the mass of the core at the time they reach their current masses (open triangles and diamonds) are in agreement with theoretical estimations.}
\label{fig:fig4} 
\end{figure}

We calculated then the simultaneous formation of the four planets for
$r_p^{min}= 20, 35$ and 40~m. Results are summarized in Table~{\ref{table:tabla4}} and Figure~\ref{fig:fig4}. Note that for $r_p^{min}= 20$~m the formation of the system changed drastically. The rapid formation of the outermost planets (especially the rapid formation of Saturn) significantly favored the formation of Jupiter. Similar results were found in our previous work (Guilera et al., 2010), in the cases where Saturn was formed before Jupiter. This rapid formation of the outermost planets induced an inner density wave responsible for the reduction in the time of formation of Jupiter and Saturn. The density wave (see Fig.~\ref{fig:fig5-2} and Fig.~\ref{fig:fig5-3}) is a product of the excitation of the planets over the planetesimals. The gravitational stirring of the planets increases planetesimal's eccentricity and inclination, and force the inner migration of planetesimals (see Eq.~\ref{eq:vel-migra}), enhancing the solids surface density and, this way, speeding up the formation of the cores of the planets. We note that we only considered the simultaneous formation of four embryos. However, oligarchic growth predicts the presence of several embryos separated by $\sim 10$ mutual Hill radii. The presence of several embryos may damp the solid density wave, but on the other hand, may also favor the formation of massive cores by their mutual collisions, which may lead to their fusion.     

For $r_p^{min}= 35$ and 40~m, the formation of the system was
quantitatively different. Although the formation time of Saturn was increased by the presence of the other embryos, it could still be considered fast. The rapid formation of  Saturn significantly decreased the formation time of Jupiter. As we can see in Fig.~\ref{fig:fig5-2} and Fig.~\ref{fig:fig5-3}, the quick formation of Saturn forced the migration of planetesimals and augmented the surface density of solids in Jupiter's neighborhood. After 0.5~Myr. the surface density of solids in Jupiter's neighborhood was increased for small planetesimals (Fig.~\ref{fig:fig5-2}) due to the presence of Saturn. For bigger planetesimals, the shape of the density wave is more evident, and in some cases the value of the surface density of solids become greater than the initial ones (Fig.~\ref{fig:fig5-3}).

In these two simulations, the cross-over times and cross-over masses of Jupiter and Saturn were in good agreement with observation and theoretical estimations of protoplanetary disks lifetimes and current solids content in the interior of the planets respectively (Fig.~\ref{fig:fig4}). Moreover, the time-scale at which both planets achieved the gaseous runaway were very similar. Regarding Uranus and Neptune, their cross-over masses were greater than their current total masses. However, for both planets, the core masses at the time they reached their current masses were in agreement with theoretical estimations. We remark that the formation time-scale of the gas giants was shorter than the formation time-scale of the ice giant.

\subsection{The case of $\Sigma \propto a^{-1}$}
\label{sec:sec4.2}

In this section, we repeat the previous methodology but now employing a surface
density profile for the disk of the form $\Sigma \propto a^{-1}$. Then, the disk
is defined by
\begin{eqnarray}
\Sigma_s(a) &=&  \left\lbrace 15.125 + \left( 60.5 - 15.125 \right)
\left[\frac{1}{2}\tanh\left(\frac{a-2.7}{0.5}\right) + 
\frac{1}{2}\right]\right\rbrace \times \nonumber \\
&&\left( \frac{a}{1~\mathrm{AU}}\right) ^{-1}~\mathrm{g~cm^{-2}},  \\
\Sigma_g(a) &=& 3954.25 \left(\frac{a}{1~\mathrm{AU}}
\right)^{-1}~\mathrm{g~cm^{-2}}.
\end{eqnarray}  
The initial mass of the disk is $\sim 0.082~\mathrm{M_{\odot}}$.

\begin{figure} 
\centering 
\includegraphics[width= 0.35\textwidth, angle= 270]{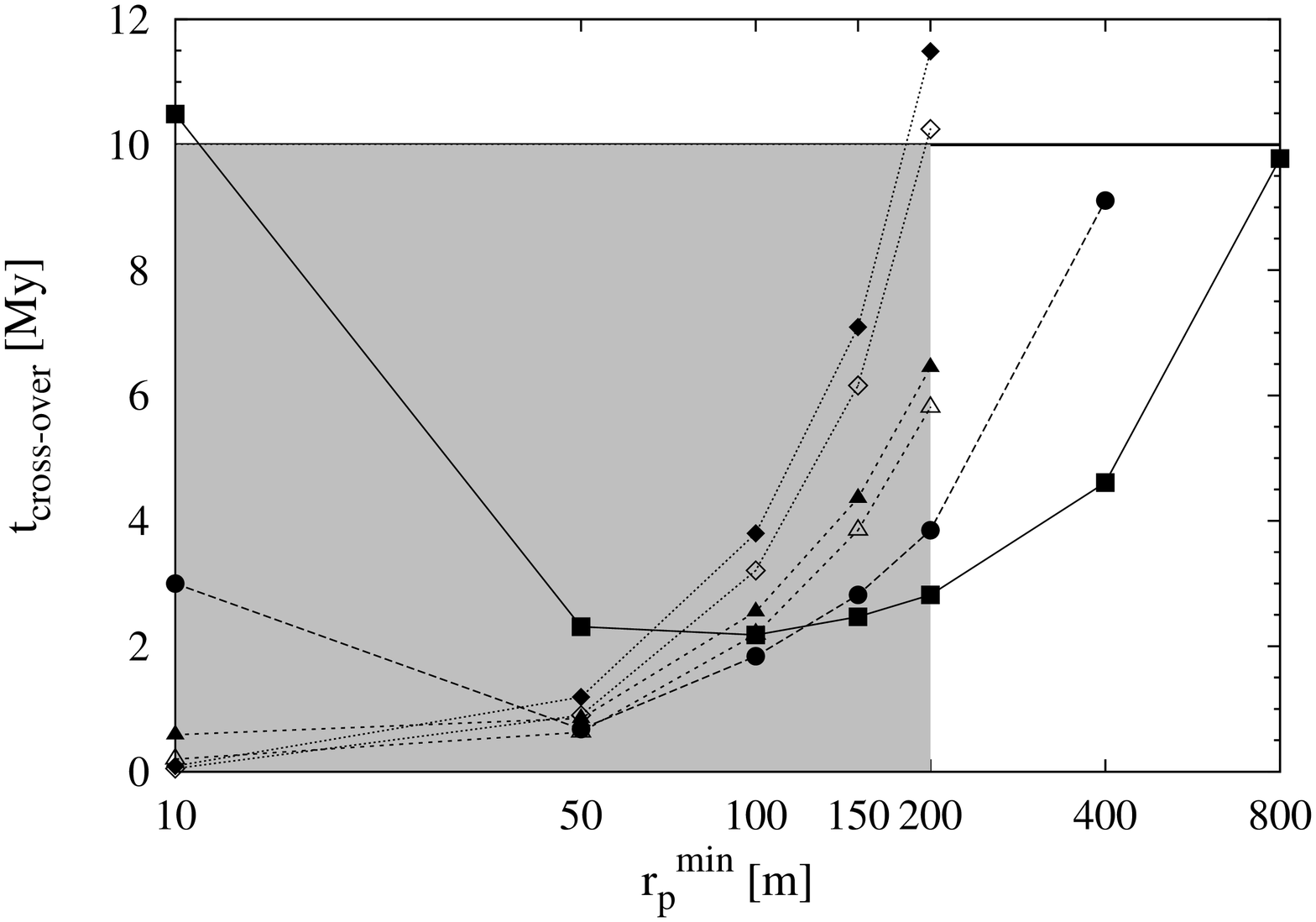} 
\caption{Same as Fig.~\ref{fig:fig3} but for a disk with a surface density
profile $\Sigma \propto a^{-1}$.}
\label{fig:fig6} 
\end{figure}

In Table~\ref{table:tabla5} and Fig.~\ref{fig:fig6} we show the results corresponding to the isolated formation of each planet. Jupiter reached its cross-over mass in less than 10~Myr for $10~\mathrm{m} \lesssim r_p^{min} \lesssim 800~\mathrm{m}$, Saturn for $10~\mathrm{m} \lesssim r_p^{min} \lesssim 400~\mathrm{m}$, while Neptune and Uranus reached their cross-over mass in less than 10~Myr for $10~\mathrm{m} \lesssim r_p^{min} \lesssim 200~\mathrm{m}$. Then, for $r_p^{min}$ between 10~m and 200~m the isolated formation of the four planets occurred in less than 10~Myr.

According to these results we calculated the simultaneous formation of the giant
planets for a size distribution of planetesimals between $r_p^{min}$
and 100~km, where we adopted for $r_p^{min}$ several discrete values between
10~m and 200~m. 

For the size distributions of planetesimals between $10~\mathrm{m} - 100~\mathrm{km}$ and $50~\mathrm{m} - 100~\mathrm{km}$, the rapid formation of the outer planets significantly augmented the planetesimals accretion rate of Jupiter and Saturn. Planetesimals accretion rate became so high ($10^{-3} - 10^{-2}~\mathrm{M_{\oplus}~yr^{-1}}$) that models corresponding to Jupiter and Saturn did not converge. We note that the mass of the cores corresponding to Jupiter and Saturn, before models did not converge, were larger than the total heavy element corresponding to Jupiter ($\sim 40~\mathrm{M_{\oplus}}$) and Saturn ($\sim 30~\mathrm{M_{\oplus}}$), respectively.

\begin{figure} 
\centering 
\includegraphics[width= 0.35\textwidth, angle= 270]{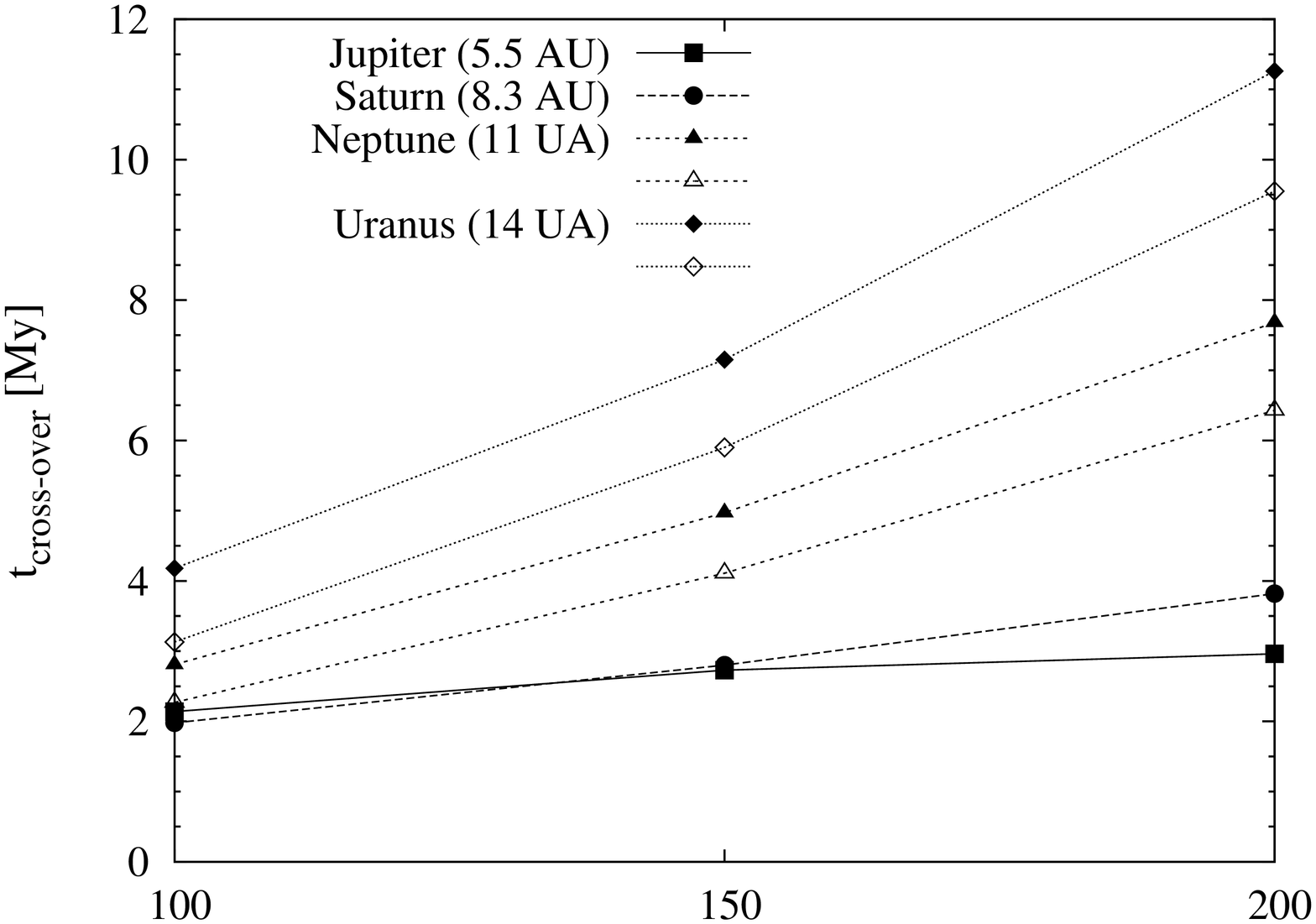}
\includegraphics[width= 0.35\textwidth, angle= 270]{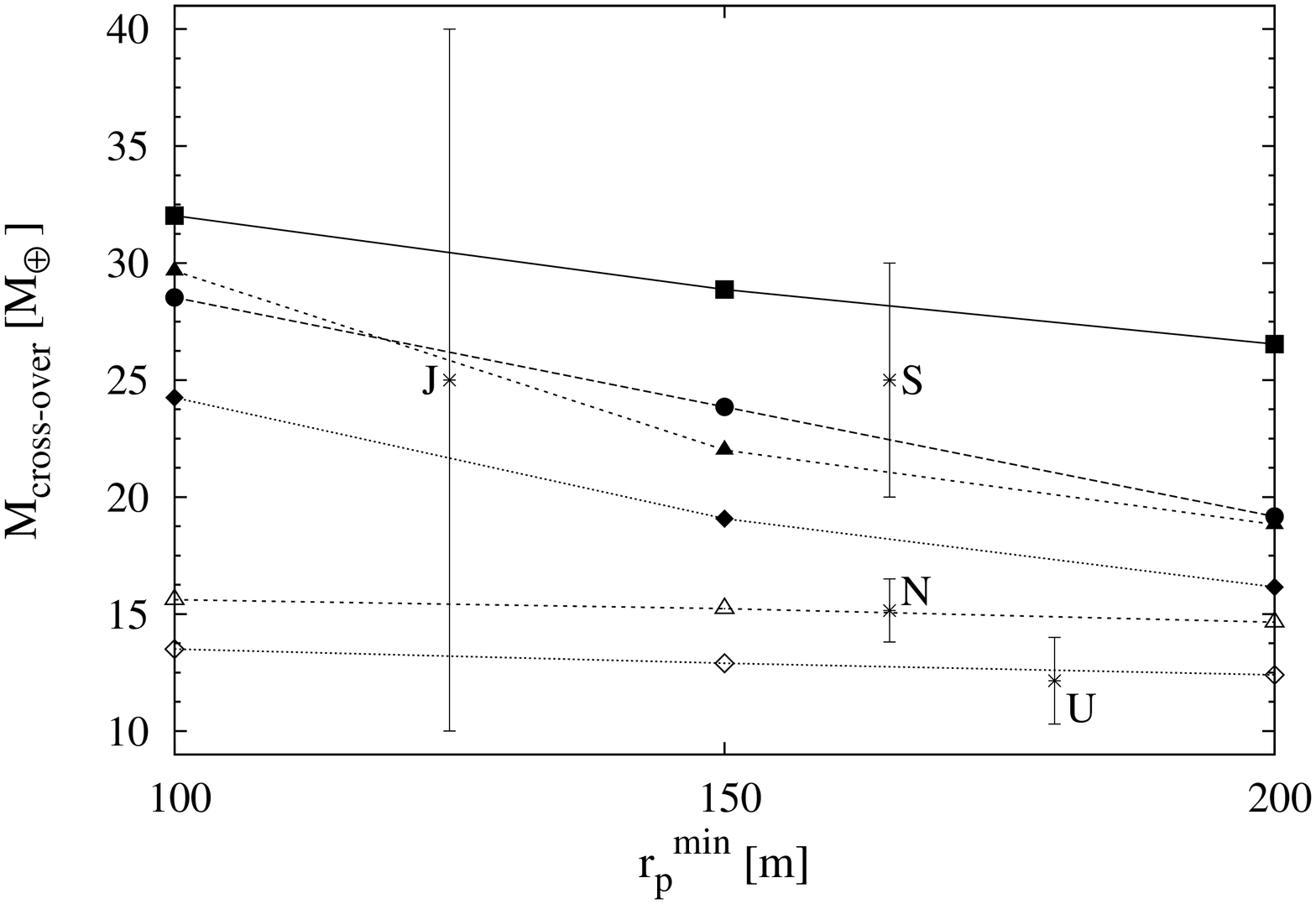}
\caption{Cross-over time (top) and cross-over mass (bottom) as function of the minimum radius of the size distribution of planetesimals for the case of the simultaneous formation. Here the disk surface density is characterized by $\Sigma \propto a^{-1}$. For all cases the cross-over times and cross-over masses of Jupiter and Saturn are in good agreement with the observational and theoretical estimations. Regarding Neptune and Uranus their cross-over masses are higher than their current total masses. However, for both planets the mass of the core at the time they reach their current masses (open triangles and diamonds) are in agreement with theoretical estimations.}
\label{fig:fig7} 
\end{figure}

On the other hand, for $r_p^{min}= 100, 150$ and 200~m, the results for the simultaneous formation were not very different from the isolated ones (see Table~\ref{table:tabla6}). We remark that for these cases our results for both the cross-over times and cross-over masses of Jupiter and Saturn were in nice agreement with the observations and theoretical estimations, respectively (Fig.~\ref{fig:fig7}). Again, the cross-over masses of Uranus and Neptune were larger than their current masses, but the core masses at the time they reached their current masses were in agreement with theoretical estimations. We note again the important fact that the time-scale of formation of the gas giants was shorter than the corresponding one to the ice giants.     

\begin{centering}
\begin{table}
\caption{Isolated formation of the solar system giant planets' cores for a disk
 with a surface density of solids and gas $\propto a^{-1}$.}
\begin{tabular}{p{0.4cm}|p{0.58cm} p{0.58cm}|p{0.58cm} p{0.65cm}|p{0.58cm}
p{0.58cm}|p{0.58cm} p{0.65cm}}
\toprule[0.8mm]

 & Jupiter & & Saturn & & Neptune & & Uranus & \\

$r_p^{min}$& $M_{over}^{cross}$ & $t_{over}^{cross}$ & $M_{over}^{cross}$& $t_{over}^{cross}$ & $M_{over}^{cross}$ & $t_{over}^{cross}$ & $M_{over}^{cross}$ & $t_{over}^{cross}$ \\
$[m]$ & $[\mathrm{M_{\oplus}}]$ & $[Myr]$ & $[\mathrm{M_{\oplus}}]$ & $[Myr]$ &
$[\mathrm{M_{\oplus}]}$ & $[Myr]$ & $[\mathrm{M_{\oplus}}]$ & $[Myr]$ \\

 \midrule[0.6mm]

10 & 24.66 & 10.49 & 25.38 & 3.00 &  30.71  &  0.59  &  42.56  &  0.09  \\  

   &       &       &       &      & (15.48) & (0.20) & (14.20) & (0.05) \\  

\midrule[0.3mm]  

50 & 26.32 & 2.31 & 34.41 & 0.68 &  32.81  &  0.85  &  30.64  &  1.19  \\

   &       &      &       &      & (15.75) & (0.63) & (13.61) & (0.90) \\

\midrule[0.3mm]

100 & 27.46 & 2.18 & 28.39 & 1.84 &  26.20  &  2.55  &  23.55  &  3.80  \\

    &       &      &       &      & (15.31) & (2.19) & (13.17) & (3.21) \\

\midrule[0.3mm] 

150 & 25.99 & 2.47 & 25.16 & 2.82 &  22.51  &  4.36  &  19.52  &  7.09  \\

    &       &      &       &      & (14.87) & (3.85) & (12.72) & (6.16) \\

\midrule[0.3mm] 

200 & 25.00 & 2.82 & 22.94 & 3.85 &  19.85  &  6.45  &  16.48  &  11.49  \\

    &       &      &       &      & (14.40) & (5.81) & (12.20) & (10.25) \\

\midrule[0.3mm] 

400 & 21.06 & 4.61 & 17.28 & 9.11 & --- & --- & --- & --- \\

\midrule[0.3mm] 

800 & 16.16 & 9.78 & --- & --- & --- & --- & --- & --- \\

\bottomrule[0.8mm] 
\end{tabular}
\label{table:tabla5} 
\end{table} 
\end{centering}

\begin{centering}
\begin{table}
\caption{Same as Table~\ref{table:tabla5} but for the simultaneous formation of
the giant planets.}
\begin{tabular}{p{0.4cm}|p{0.58cm} p{0.58cm}|p{0.58cm} p{0.65cm}|p{0.58cm}
p{0.58cm}|p{0.58cm} p{0.65cm}}
\toprule[0.8mm]

 & Jupiter & & Saturn & & Neptune & & Uranus & \\
 $r_p^{min}$& $M_c$ & $t_f$ & $M_c$& $t_f$ & $M_c$ & $t_f$ & $M_c$ & $t_f$ \\
$[m]$ & $[\mathrm{M_{\oplus}}]$ & $[Myr]$ & $[\mathrm{M_{\oplus}}]$ & $[Myr]$ &
$[\mathrm{M_{\oplus}}]$ & $[Myr]$ & $[\mathrm{M_{\oplus}}]$ & $[Myr]$ \\

\midrule[0.6mm]

100 & 32.03 & 2.14 & 28.53 & 1.98 &  26.97  &  2.81  &  24.25  &  4.18  \\

    &       &      &       &      & (15.61) & (2.27) & (13.50) & (3.13) \\

\midrule[0.3mm] 

150 & 28.87 & 2.73 & 23.85 & 2.80 &  22.01  &  4.97  &  19.08  &  7.15 \\

    &       &      &       &      & (15.23) & (4.11) & (12.90) & (5.90) \\

\midrule[0.3mm] 

200 & 26.54 & 2.96 & 19.17 & 3.82 &  18.83  &  7.68  &  16.15  &  11.26  \\

    &       &      &       &      & (14.65) & (6.43) & (12.40) &  (9.55) \\

\bottomrule[0.8mm] 
\end{tabular}
\label{table:tabla6} 
\end{table} 
\end{centering}

\subsection{The case of $\Sigma \propto a^{-1/2}$}
\label{sec:sec4.3} 

Finally, we repeated the calculations employing a very smooth surface density
profile, proportional to $a^{-1/2}$. In this case, the resulting expressions for
$\Sigma$
are:

\begin{eqnarray}
\Sigma_s(a) &=&  \left\lbrace 6.45 + \left( 25.8 - 6.45 \right)
\left[\frac{1}{2}\tanh\left(\frac{a-2.7}{0.5}\right) + 
\frac{1}{2}\right]\right\rbrace \times \nonumber \\
&&\left( \frac{a}{1~\mathrm{AU}}\right) ^{-1/2}~\mathrm{g~cm^{-2}},  \\
\Sigma_g(a) &=& 1686.25 \left(\frac{a}{1~\mathrm{AU}}
\right)^{-1/2}~\mathrm{g~cm^{-2}}.
\end{eqnarray}  
The initial mass of the disk in this case was $\sim 0.13~\mathrm{M_{\odot}}$.

\begin{figure} 
\centering 
\includegraphics[width= 0.345\textwidth, angle= 270]{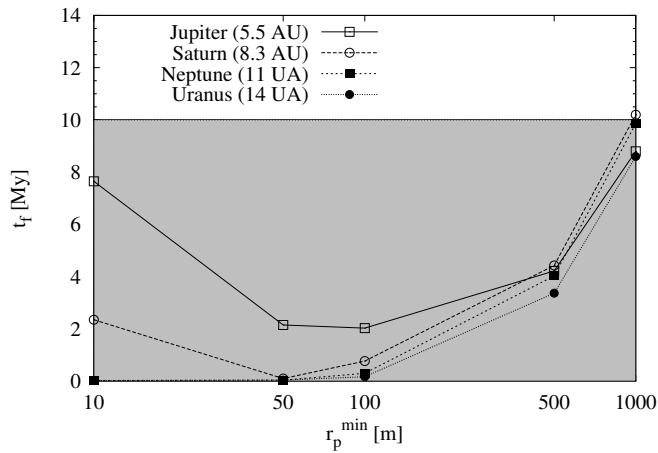} 
\caption{Same as Fig.~\ref{fig:fig3} and Fig.~\ref{fig:fig6} but for a disk with
a
surface density profile $\Sigma \propto a^{-1/2}$.}
\label{fig:fig8} 
\end{figure}

In Table~\ref{table:tabla7} we show the results corresponding to the
isolated formation of each planet. We see that each planet reached their cross-over mass in less than $\sim 10$~Myr with a size distribution of planetesimals ranging from $r_p^{min}$ to $100~\mathrm{km}$, where $r_p^{min}$ could take values between 10~m and 1~km (Fig.~\ref{fig:fig8}). However, for this density profile, there were cases where the final solid masses in the planets interior were higher than the theoretical estimations. 

The results of the simultaneous formation for this profile are summarized in
Table~\ref{table:tabla8}. 

For $r_p^{min}$ between 10 and 100~m, the fast formation of Neptune and Uranus significantly augmented the planetesimals accretion rate of Jupiter and Saturn. Planetesimals accretion rate became so high ($10^{-3} - 10^{-2}~\mathrm{M_{\oplus}~yr^{-1}}$) that models corresponding to Jupiter and Saturn did not converge. We note again that the mass of the cores corresponding to Jupiter and Saturn, before models did not converge, were larger than the total heavy element corresponding to Jupiter ($\sim 40~\mathrm{M_{\oplus}}$) and Saturn ($\sim 30~\mathrm{M_{\oplus}}$), respectively. For $r_p^{min}= 500$~m and $r_p^{min}= 1$~km the presence of Uranus reduced the formation time of the inner planets by a factor of 2, but increased significantly its final solid masses. While the cross-over time of Jupiter and Saturn are in nice agreement with theoretical estimations, the cross-over mass corresponding to Jupiter and Saturn are larger than the total heavy elements content corresponding to Jupiter ($\sim 40~\mathrm{M_{\oplus}}$) and Saturn ($\sim 30~\mathrm{M_{\oplus}}$). 

\begin{centering}
\begin{table}
\caption{Isolated formation of the giant planets for a disk
with a surface density of solids and gas $\propto a^{-1/2}$. 
}
\begin{tabular}{p{0.45cm}|p{0.58cm} p{0.58cm}|p{0.58cm} p{0.65cm}|p{0.58cm}
p{0.58cm}|p{0.58cm} p{0.6cm}}
\toprule[0.8mm]

 & Jupiter & & Saturn & & Neptune & & Uranus & \\
 $r_p^{min}$& $M_c$ & $t_f$ & $M_c$& $t_f$ & $M_c$ & $t_f$ & $M_c$ & $t_f$ \\
$[m]$ & $[\mathrm{M_{\oplus}}]$ & $[Myr]$ & $[\mathrm{M_{\oplus}}]$ & $[Myr]$ &
$[M_{\oplus}]$ & $[Myr]$ & $[\mathrm{M_{\oplus}}]$ & $[Myr]$ \\

 \midrule[0.6mm]

10 & 25.27 & 7.62 & 26.00 & 2.33 &  34.55  &  0.09  &  59.85  &  0.03  \\  

   &       &      &       &      & (16.50) & (0.03) & (14.30) & (0.02) \\

\midrule[0.3mm]  

50 & 28.65 & 2.14 & 42.00 & 0.10 &  54.15  &  0.06  &  55.50  &  0.06  \\

   &       &      &       &      & (16.52) & (0.04) & (14.20) & (0.04) \\

\midrule[0.3mm] 

100 & 31.31 & 2.02 & 35.00 & 0.77 &  36.73  &  0.44  &  38.21  &  0.30 \\

    &       &      &       &      & (15.99) & (0.30) & (13.95) & (0.17) \\

\midrule[0.3mm] 

500 & 23.86 & 4.19 & 23.65 & 4.42 &  23.32  &  4.53  &  23.38  &  4.00  \\

    &       &      &       &      & (14.95) & (4.03) & (13.14) & (3.37) \\

\midrule[0.3mm] 

1000 & 18.64 & 8.73 & 17.88 & 10.14 &  17.57  & 10.65  &  17.83  &  9.72 \\

     &       &      &       &       & (13.81) & (9.86) & (12.42) & (8.60) \\

\bottomrule[0.8mm] 
\end{tabular}
\label{table:tabla7} 
\end{table} 
\end{centering}

\begin{centering}
\begin{table}
\caption{Same as Table~\ref{table:tabla7} but for the simultaneous formation.}
\begin{tabular}{p{0.45cm}|p{0.58cm} p{0.58cm}|p{0.58cm} p{0.65cm}|p{0.58cm}
p{0.58cm}|p{0.58cm} p{0.6cm}}
\toprule[0.8mm]

 & Jupiter & & Saturn & & Neptune & & Uranus & \\
 $r_p^{min}$& $M_c$ & $t_f$ & $M_c$& $t_f$ & $M_c$ & $t_f$ & $M_c$ & $t_f$ \\
$[m]$ & $[\mathrm{M_{\oplus}}]$ & $[Myr]$ & $[\mathrm{M_{\oplus}}]$ & $[Myr]$ &
$[\mathrm{M_{\oplus}}]$ & $[Myr]$ & $[\mathrm{M_{\oplus}}]$ & $[Myr]$ \\

\midrule[0.6mm]

500 & 53.37 & 2.42 & 35.54 & 2.37 &  22.62  &  2.92  &  23.20  &  4.50  \\

    &       &      &       &      & (15.67) & (2.29) & (13.04) & (3.78) \\

\midrule[0.3mm] 

1000 & 47.90 & 4.80 & 31.05 & 4.73 &  16.85  &  6.07  &  16.21  &  11.50 \\

     &       &      &       &      & (13.87) & (5.06) & (12.24) & (10.09) \\

\bottomrule[0.8mm] 
\end{tabular}
\label{table:tabla8} 
\end{table} 
\end{centering}

\section{Discussion and conclusions}

In this paper we studied the {\it in situ}, simultaneous formation of the Solar
System giant planets in the framework of the core accretion model, and according to the conditions imposed by the Nice model. The goal of this work was to delimit some parameters that describe the protoplanetary disk (especially the surface density profile and planetesimal's population size distribution) in relation to the likelihood of the formation of the four giant planet's cores, in less than $\sim 10$~Myr and with a content of heavy elements in good agreement with current estimations.

The surface density profile of the disk was assumed as a power law, $\Sigma \propto a^{-p}$, where the power index $p$ is considered as a free parameter, and several values of $p$ were analyzed. We considered that accreted planetesimals follow a power law mass distribution of the form  $n(m) \propto m^{-2.5}$, where most of the mass is in smaller objects. We assumed planetesimals to be spheres of constant density, which in turn implies a power law distribution for planetesimal's radii. We discretized the continuous planetesimal population, considering that the minimum radius,$r_p^{min}$, is also a free parameter and the maximum radius is fixed in $r_p^{max}=100$~km. 

\begin{figure} 
\centering 
\includegraphics[width= 0.35\textwidth, angle= 270]{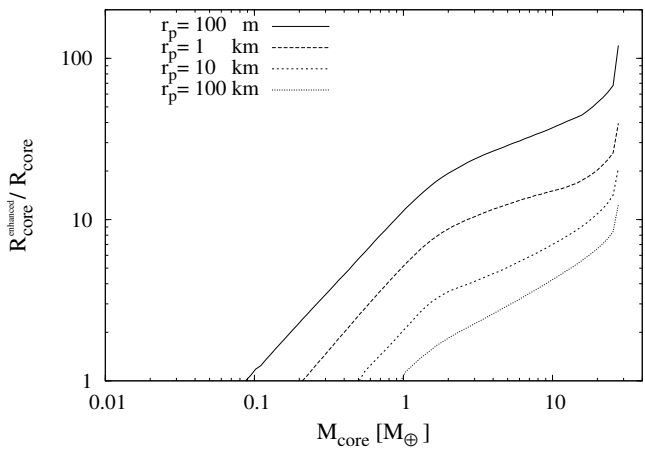} 
\caption{The ratio between the enhanced radius and the core radius as function of the core mass for the isolated formation of Jupiter, for a disk with a density profile $\propto a^{-1}$, and with a planetesimal size distribution between 100~m and 100~km. The envelope gas drag for smaller planetesimals becomes effective at very small core masses.}
\label{fig:rr} 
\end{figure}

\begin{figure}[]
\centering 
\includegraphics[width= 0.34\textwidth, angle= 270]{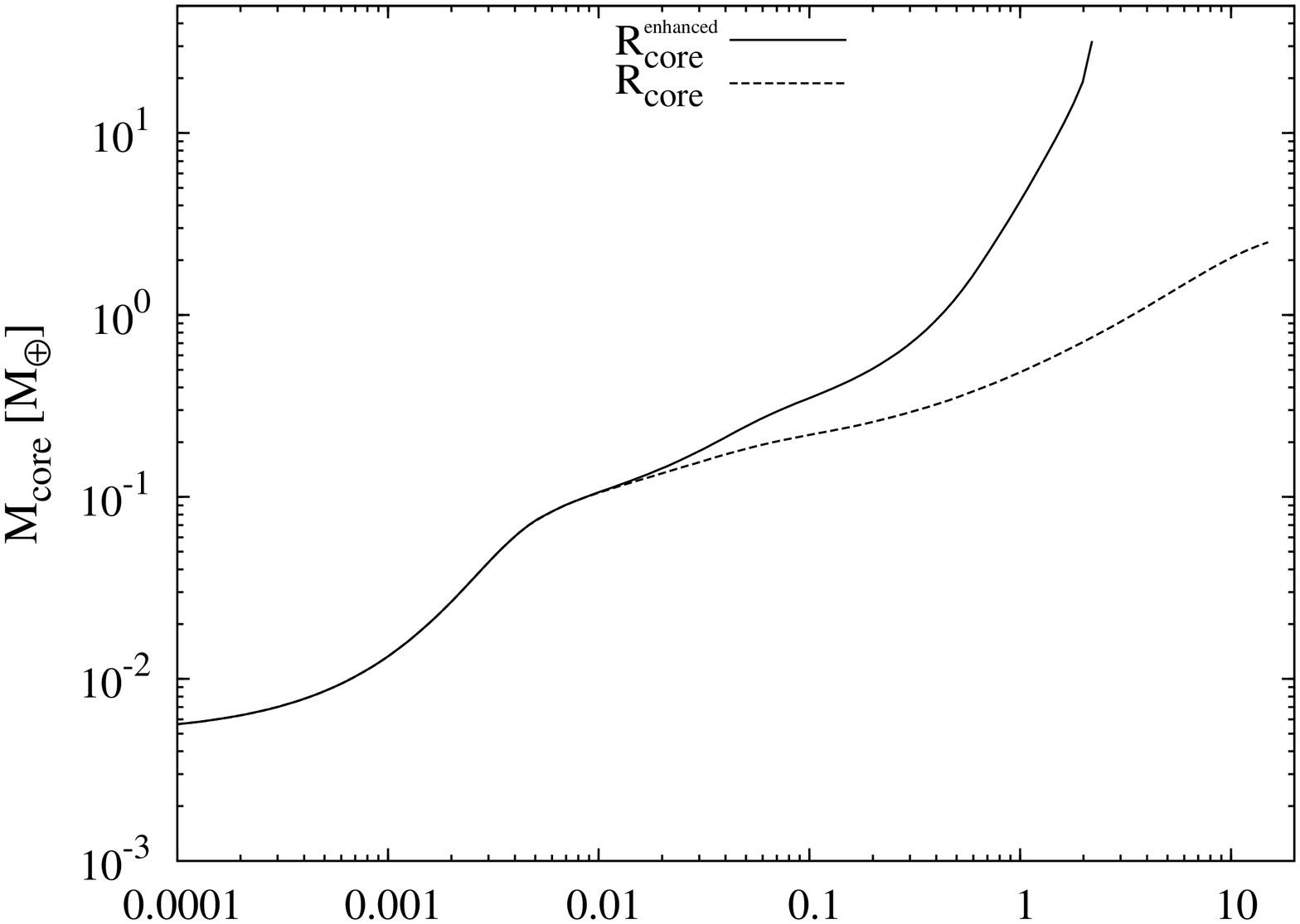}
\includegraphics[width= 0.34\textwidth, angle= 270]{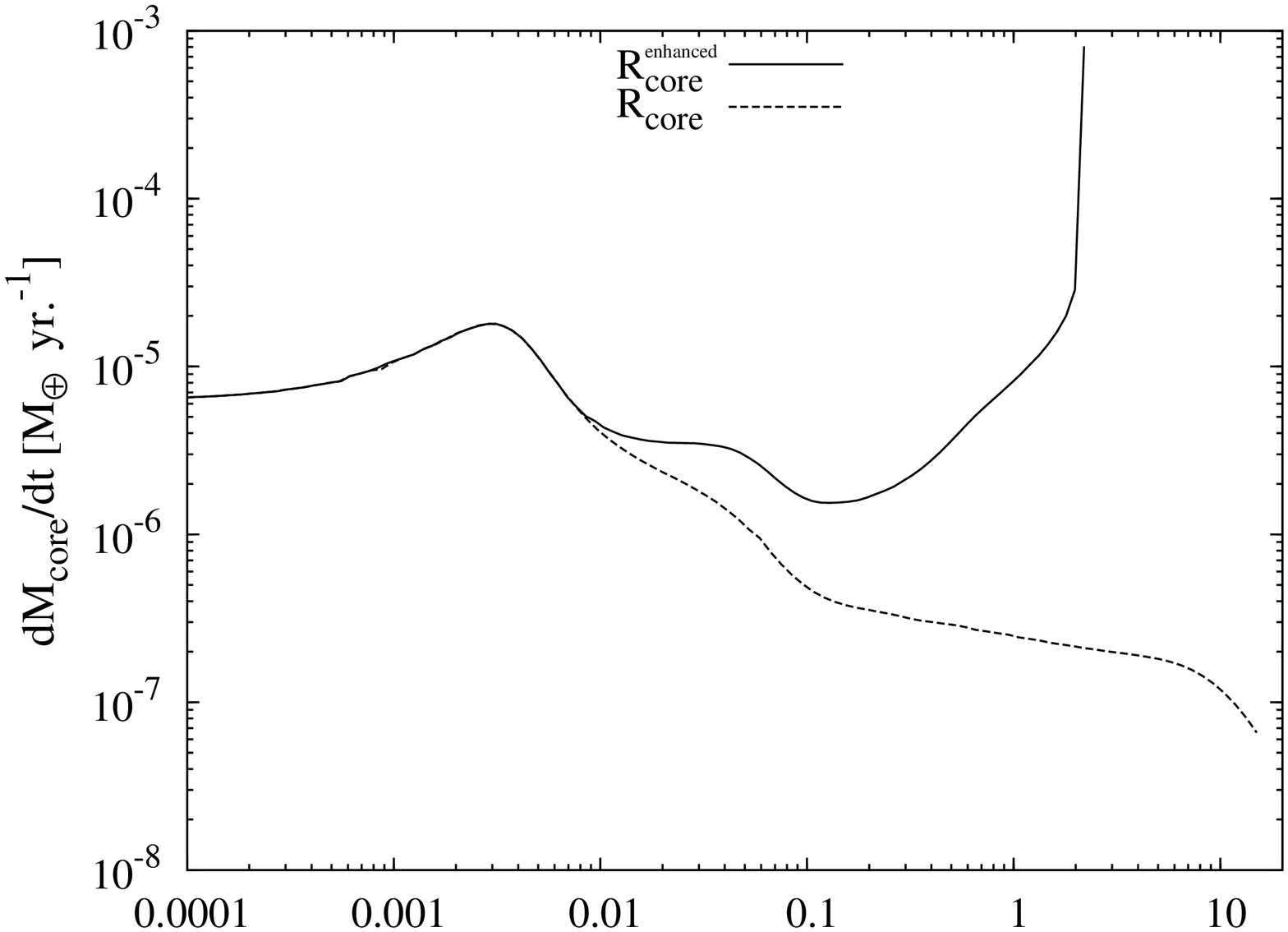}
\includegraphics[width= 0.34\textwidth, angle= 270]{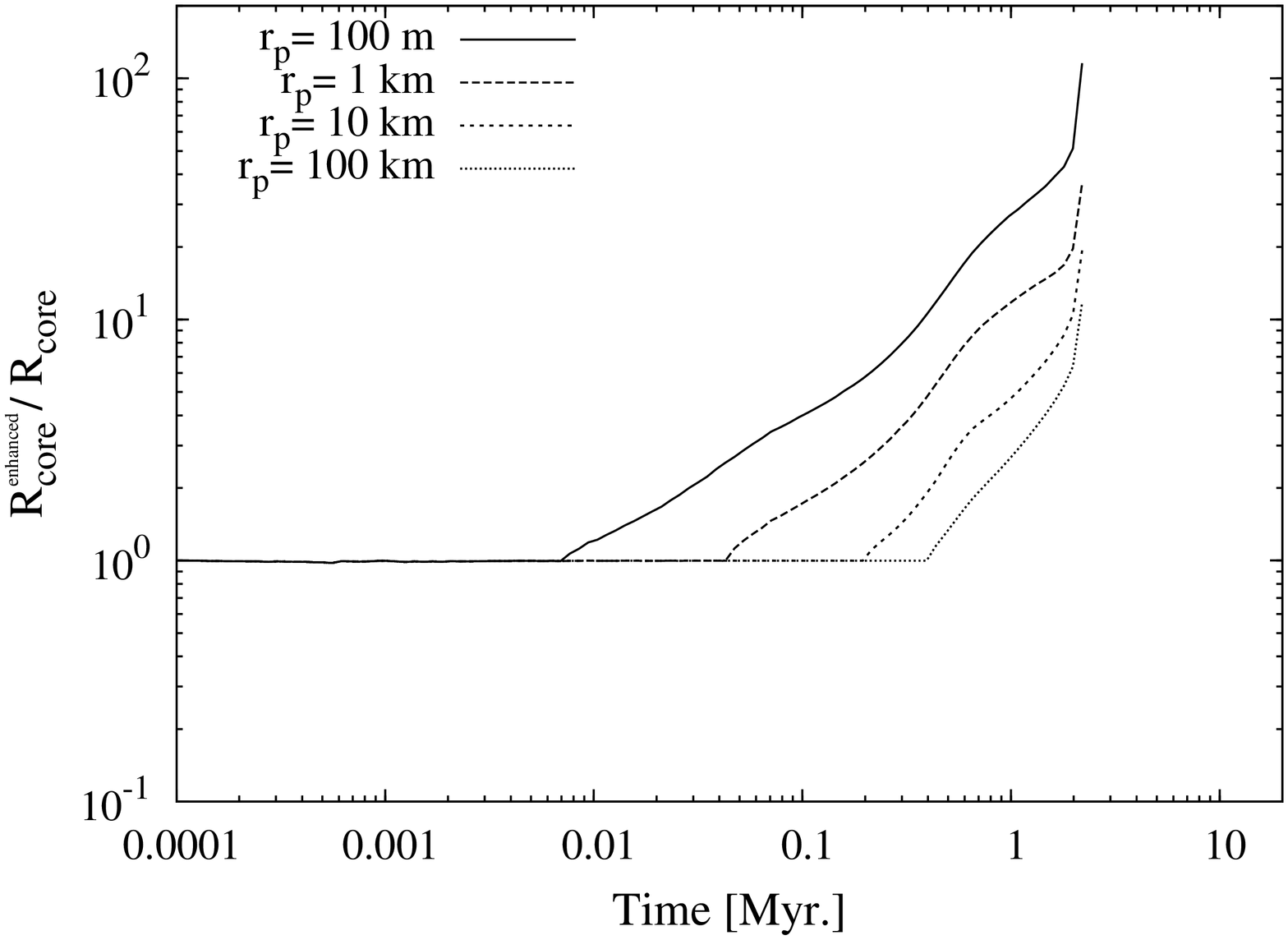}
\caption{Top and middle: Time evolution of the core mass and the planetesimal accretion rate for the isolated formation of Jupiter employing a disk with density profiles $\propto a^{-1}$, and a planetesimal size distribution between 100~m and 100~km, with (solid line) and without (dashed line) considering the enhancement of the capture cross-section due to the envelope gas drag.
Bottom: The ratio between the enhanced radius and the core radius as function of time for the case where the enhancement of the capture cross-section is incorporated.}
\label{fig:jup-comp} 
\end{figure}

We note that, although we calculate the evolution of the
planetesimal disk due to planet's accretion and planetesimal migration,
 we do not calculate the planetesimal's evolution due
to aggregation and fragmentation. Inaba et al.~(2003) found that large
amounts of mass could be lost due to inward drift of small collision
fragments. So, fragmentation seems to play an important role and it should be considered
in a more accurate model. Although we can not predict how fragmentation
could affect the results of our model, we want to point out some differences
between the working hypothesis of Inaba et al.~(2003) and the one
presented here. Inaba et al.~(2003) started with an homogeneous
population of planetesimals of radii $\sim 10$~km. So, smaller
planetesimals appear due to fragmentation of bigger ones. This implies
that most of the mass relies in bigger planetesimals (Dohnanyi, 1969; Wetherill \&
Stewart, 1993). On the other hand, we started our simulations already in the
oligarchic growth regime, with an embryo with the mass of the Moon
($\sim 0.01~\mathrm{M_{\oplus}}$) immersed in a swarm of planetesimals
that follow a mass distribution. Ormel et al.~(2010) employing statistical
simulations that include several physical processes such as dynamical
friction, viscous stirring, gas drag, and fragmentation, found that
starting with an homogeneous populations of planetesimals of radii $r_0$, the transition between the runaway growth and oligarchic growth
is characterized by a power-law size distribution of mass index
$q\sim2.5$, where most of the mass lie in small planetesimals. So, our
initial conditions are consistent with the oligarchic growth regime
using $r_0 \lesssim 1$~km. Another important difference is the moment at which the gas drag of the envelope become efficient. Inaba et al.~(2003) found that when the mass of the embryo reach $\sim$ the mass of Mars it is capable to acquire an envelope. However, we found that when the mass of the embryos is $\sim 0.1~\mathrm{M_{\oplus}}$ it already has an envelope and the envelope gas drag becomes efficient for smaller planetesimals (tens and hundreds meter-size planetesimals). Moreover, as we can see in Fig.~\ref{fig:rr} when the mass of the embryo is $\sim 0.6~\mathrm{M_{\oplus}}$ the ratio between the \textit{enhanced radius} and the core radius is a factor $\sim 7$. The enhanced of the cross-section capture due to the presence of the envelope and the moment when it become efficient are very important. As an example, we calculated the isolated formation of Jupiter for a disk with density profiles $\propto a^{-1}$, and with a planetesimal size distribution between 100~m and 100~km, but now not taking into account the enhancement of the capture cross-section  due to the envelope gas drag. We compared this simulation to the one with the same initial conditions but considering the enhancement of the capture cross-section. In Fig.~\ref{fig:jup-comp} we show the comparison between the two simulations. For the case where the enhanced capture cross-section is considered, Jupiter reached its cross-over mass at 2.18~Myr. (the value of the cross-over mass is $27.46~\mathrm{M_{\oplus}}$, see Table~\ref{table:tabla5}), while if the enhanced capture cross-section is not considered, Jupiter did not reach its cross-over mass (the simulation was halted at 15~Myr. and Jupiter only reached a core of $\sim 2.5~\mathrm{M_{\oplus}}$ with a negligible envelope). We see that when the core -for the case where the enhancement of the capture cross-section is not incorporated- reached the mass of Mars, the corresponding core for the case where the enhanced capture cross-section is considered, is able to grow up to  $10~\mathrm{M_{\oplus}}$ in the same elapsed time. It is clear that the enhancement of the cross-section of the core due to the presence of the envelope plays a fundamental role in the formation of a giant planet. We also can see how the planetesimal accretion rate suffers a significant drop when the enhancement of the cross-section is not taken into account. Finally, we can see that the difference between both simulation begins when the envelope gas drag become effective for small planetesimals (at $\sim 0.01$~Myr, which correspond to a core mass of $\sim 0.1~\mathrm{M_{\oplus}}$). So, the moment (or the mass of the embryo) at which the envelope gas drag becomes effective for smaller planetesimals is important due to the fact that we are considering that most of the mass lies in small objects. The difference in the ratio between the enhanced radius and the core radius, respect to Inaba et al.~(2003), may correspond to the fact that we are employing a more accurate model to calculate the growth and evolution of the planet's envelope. Therefore, we conclude that fragmentation could be an important effect, and we plan to include it in future improvements of our model, but we think that a definitive evaluation of its relevance should await calculations including a realistic model of the process, combined with a detailed prescription of the effect of the planet atmosphere on the fragments.

We started studying a Desch-like profile for the disk surface density, $\Sigma
\propto a^{-2}$ (see Desch, 2007).
 We first computed the isolated formation of the four giant planets looking for
a common interval for the minimum radius of the planetesimal population that could allow the simultaneous
formation of all the planets. However, we did not find overlapping values for
the four planets. Despite that, we anyway calculated some simulations in order
to check if the simultaneous formation was possible (see Table
\ref{table:tabla2}), and confirmed that, according to our model, adopting
Desch profile makes the formation of the giant planets very unlikely, as the formation time-scale largely exceeds the 10 Myr barrier.

We proceeded then with less steep profiles. We analyzed other three cases:
 $\Sigma \propto a^{-1.5}$, $\Sigma \propto a^{-1}$ and $\Sigma \propto
a^{-0.5}$, 
the first one corresponding to the standard minimum mass solar nebula and the
other two following results of  Andrews et al. (2009; 2010), which are based
on observational data of
circumstellar disks. In all these cases, when first calculating the isolated
formation of Jupiter, Saturn, Neptune and 
Uranus, we were able to find several common values for $r_p^{min}$ that
could lead to a satisfactory result when calculating the 
simultaneous formation (see Figs.~\ref{fig:fig3},
~\ref{fig:fig6},~\ref{fig:fig8}). Indeed, for these
values we run the corresponding simulations for the simultaneous formation. For
these three density profiles, the simultaneous formation could be, in principle,
considered as possible if the most abundant planetesimals are those whose radii
range from several tens to several hundreds of meters depending on the
case, but not larger than that. 
 From our results it is clear that the 
smoother the profile density of the disk, the more efficient the formation
process.
Moreover, considering lower values for $p$ allowed us to increase the minimum
radius of the accreted
planetesimals. 
We note that decreasing $p$ is directly related to considering more massive
nebulae, which in turn means that the mass in the protoplanets feeding zone is
larger, favoring this way the accretion process. Furthermore, planetesimal
migration is less efficient because gas drag is weaker, so planetesimals remain
longer in the feeding zone of the protoplanets. Beside this, for a solid surface
density profile $\Sigma_s \propto a^{-p}$ where $p < 1$ the mass of solids grows
outwards the planetesimal disk, and the incoming mass flux in the feeding
zone of the protoplanets is greater than the out-coming one, which also favors
the accretion.

On the other hand, Desch's profile is compatible with a
decretion disk. This kind of profiles turned out to be almost stationary for
about ten million
years. Crida (2009) demonstrated that in a disk with this density profile, the
four
giant planets of the solar system would have been unable to survive.
Particularly, Jupiter would have become a hot giant planet. Morbidelli \& Crida
(2007) and Crida (2009)  also
showed that Jupiter and Saturn could avoid migration if a less massive nebula,
like Hayashi's nebula is
assumed.
While the initial value of the gas density used in our simulations, at the
position of Jupiter, is about 5 times higher than that of Hayashi's minimum mass
solar nebula,
this value is decreased by an exponential dissipation factor during the
formation of the planets. Then, by the time 
Jupiter and Saturn finished their formation, the disk conditions correspond
to a much less massive disk, especially in the cases where the formation times of
Jupiter
and Saturn are similar to the dissipation time-scale of the gas disk. This
should help to prevent Jupiter and Saturn migration in initially massive
protoplanetary disks.

We found several values of $r_p^{min}$ for each profile considered here
(except $p=2$), for which the simultaneous formation of the planets could be achieved in less than 10~Myr and the amount of heavy elements in the
interior of the planets were in good agreement with theoretical
estimations of current abundances. In most of the cases, the formation time-scale of the gas giants was shorter than the formation time-scale of the ice
giants. However, when very small planetesimals were the most abundant
bodies in the disk, the icy giants formed before the gas giants. We consider these cases very unlikely, at least under the hypothesis of our model, as they would need extra explanations to justify why they did not started the gaseous runaway before Jupiter and Saturn if, in principle, there was still plenty of material that could be accreted. This specially applies for the cases where Uranus and/or Neptune were the first to be formed.

There is another interesting result we would like to highlight. When we consider $\Sigma \propto a^{-1}$ and $r_p^{min}=100$~m (Sec. \ref{sec:sec4.2}) the cross-over times of Jupiter and Saturn and the time at which Neptune and Uranus reached their current masses are of the same time-scale; around $2 - 3$~Myr (see Table~\ref{table:tabla6} and Fig.\ref{fig:fig7}). This time-scale is shorter than the dissipation time-scale of the disk. The fact that the four planets formed on the same time-scale, which is also comfortably short, is not a minor point, but on the contrary it presents a scenario that
should be explored in more detail in the future. If the gaseous component of the
disk could be rapidly evaporated by some external mechanism [for instance, if
the solar system is strongly irradiated by an OB stellar
association (see, for example, the work by Clarke (2007))], our results can
provide a pathway to find suitable parameters to describe the protoplanetary
disk. Then, we note that there are cases where calculations of simultaneous formation with an appropriate planetesimal size distribution, and together with a smooth surface density profile, provide suitable conditions for the timely formation of the external planets, the four formation processes occurring on the same time-scale.

Finally, we would like to remark the need that the size distribution of planetesimals extends to objects with radii $< 1$~km to form the four planets on a time-scale compatible with the observational estimates. It was already pointed out in Goldreich et al. (2004) that the presence of a large amount of small planetesimals would help to speed up planetary formation. If the planetesimal disk is dominated by big bodies the time-scale to form solid embryos able to bind a significant envelope and start the gaseous runaway accretion would be too long to complete the formation in less than 10 Myr.

However, this result contradicts recent models for planetesimal
formation. Johansen et al.~(2007), Cuzzi et al.~(2008) and Youdin~(2011) describe different mechanisms for planetesimal formation incorporating turbulence as a way to aid planetesimals to grow, showing that, if planetesimals formed this way, they are likely to be large ($\sim 100$~km, or even bigger, growing directly from sub-m particles). However, these models do not predict a specific planetesimal size distribution and the nature of the turbulence is controversial. Nelson \& Gressel~(2010) showed that fully developed magneto-hydrodynamic turbulence in protoplanetary discs would have a destructive effect on embedded planetesimals. Arguments in favor of the hypothesis of primitive initial large planetesimals are found in Morbidelli et al.~(2009), where through models of the collisional evolution of the asteroid belt they conclude that the primitive asteroids should have been big. However, their model does not take into account a very important effect:
the primitive intense bombardment of the asteroid belt by outer planetary region
comets (Gil Hutton \& Brunini, 1999) that should change this conclusion. Moreover, Weidenschilling~(2010) proposed an alternative scenario to the work of Morbidelli et al.~(2009). He found that coagulation from small planetesimals ($\sim 100$~m of diameter) represents better the size distribution of the asteroid belt. Such small planetesimals could be formed by coagulation in collision driven processes by size-dependent drift due to nebular gas drag (Weidenschilling, 1997). For these reasons, we consider that the planetesimal size issue is far from being fully understood, and a distribution where most of the solid mass accreted by a protoplanet comes from small planetesimals cannot be ruled out.  

However, one possible way to relax this assumption is to consider giant impacts
and mergers during the planetary formation (Li et al., 2010). If a forming giant
planet is impacted by a Mars- to an Earth-mass embryo, 
the impactor most likely reaches the core possibly remixing the core into the
envelope (Li et al., 2010). 
Broeg \& Benz (2011) have studied the effect of this impact on the gas accretion
rate: initially, most of the envelope can be ejected, but afterwards gas is
reaccreted very fast and the overall gas accretion rate turns out to be larger
than in the 
standard, continuous-planetesimal-accretion scenario. As a consequence, the
planet gains large amounts of mass in a short time-scale and the
growth-time-scale is reduced.

Another path to accelerate the formation of the gaseous planets is to consider
the fusion of embryos as a mechanism to obtain massive cores,
especially in the early stages of the disk evolution. Oligarchic growth predicts
the simultaneous formation of many embryos on orbits separated by about 10 Hill
radii from each other. Some works have been done to study the fusion of growing
protoplanets but only
taking into account their solid cores, neglecting the presence of the envelope
(Chambers, 2006; Brunini \& Benvenuto, 2008). 
To perform a full-detailed simulation in the context of the present study we
would need to 
compute the merger of embryos self-consistently, taking into account the growing
gaseous
envelope. This is a very complex phenomenon, beyond the scope of the present
paper, but that will be investigated
in a future work. 


\begin{acknowledgements}
We wish to acknowledge the useful comments of C. Broeg, A. Crida, L. Fouchet, Y. Alibert, and Y. Miguel. We also appreciate the constructive criticism that John Chambers (the referee) made about our work, which had helped us to significantly improved this work. A. F. was supported by the European Research Council under grant number 239605.
\end{acknowledgements}




\begin{thebibliography}{}

\bibitem[1976]{adachi} Adachi, I., Hayashi, C., \& Nakazawa, K.\ 1976, Progress
of Theoretical Physics, 56, 1756 

\bibitem[b]{alibert_2005} Alibert, Y., Mousis, O., Mordasini, C., \& Benz, W.\ 2005, \apjl, 626, L57

\bibitem[2009]{andrews_2009} Andrews, S.~M., Wilner, D.~J., Hughes, A.~M., Qi,
C., \& Dullemond, C.~P.\ 2009, \apj, 700, 1502

\bibitem[2010]{andrews_2010} Andrews, S.~M., Wilner, 
D.~J., Hughes, A.~M., Qi, C., \& Dullemond, C.~P.\ 2010, accepted in \apj,(print
version, arXiv:1007.5070) 

\bibitem[2010]{bat_brown_2010} Batygin, K., \& Brown, M.~E.\ 2010, \apj, 716, 1323 

\bibitem[2005]{benve_2005} Benvenuto, O.~G., \& Brunini, A.\ 2005, \mnras, 356,
1383 

\bibitem[2009]{bfb} Benvenuto, O.~G., Fortier, A., \& Brunini, A.\ 2009, Icarus,
204, 752 

\bibitem[2010]{broegbenz} Broeg, C. and Benz, W. \ 2010, in prep. 

\bibitem[2008]{bb} Brunini, A., \& Benvenuto, O.~G.\ 2008, Icarus, 194, 800 

\bibitem[2006]{chambers} Chambers, J.\ 2006, Icarus, 180, 496 

\bibitem[2010]{chambers_2010} Chambers, J.~E.\ 2010, \icarus, 208, 505 

\bibitem[2007]{clarke2007} Clarke, C.~J.\ 2007, \mnras,  376, 1350

\bibitem[2009]{crida_2009} Crida, A.\ 2009, \apj, 698, 606

\bibitem[2008]{cuzzi_2008} Cuzzi, J.~N., Hogan, R.~C., \& Shariff, K.\ 2008, \apj, 687, 1432 

\bibitem[2007]{desch_2007} Desch, S.~J.\ 2007, \apj, 671, 878 

\bibitem[1969]{dohnanyi_1969} Dohnanyi, J.~S.\ 1969, \jgr, 74, 2531 

\bibitem[2007]{fbb} Fortier, A., Benvenuto, O.~G., \& Brunini, A.\ 2007, \aap,
473, 311 

\bibitem[2009]{fbb2} Fortier, A., Benvenuto, O.~G., \& Brunini, A.\ 2009, \aap,
500, 1249

\bibitem[1999]{gh-1999} Gil-Hutton, R., \& Brunini, A.\ 1999, \planss, 47, 331 

\bibitem[2004]{gls_2004} Goldreich, P., Lithwick, Y., \& Sari, R.\ 2004, \araa, 42, 549 

\bibitem[2005]{gomes_2005} Gomes, R., Levison, 
H.~F., Tsiganis, K., \& Morbidelli, A.\ 2005, \nat, 435, 466

\bibitem[2010]{guilera_2010} Guilera, O.~M., Brunini, A., \& Benvenuto, O.~G.\ 2010, \aap, 521, A50 

\bibitem[2005]{guillot} Guillot, T.\ 2005, Annual Review of Earth and Planetary
Sciences, 33, 493 

\bibitem[2009]{guillot2} Guillot, T., \& Gautier, D.\ 2009, Treatise of
Geophysics, vol. 10, Planets and Moons, Schubert G., Spohn T. (Ed.) (2007)
439-464 (print version, arXiv:0912.2019). 

\bibitem[2001]{haisch} Haisch, K.~E., Jr., Lada, E.~A., \& Lada, C.~J.\ 2001,
\apjl, 553, L153 

\bibitem[1981]{hayashi} Hayashi, C.\ 1981, Progress of  Theoretical Physics
Supplement, 70, 35 

\bibitem[2001]{inaba_2001} Inaba, S., Tanaka, H., Nakazawa, K., Wetherill, G.~W., \& Kokubo, E.\ 2001, Icarus, 149, 235 

\bibitem[2003]{inaba_2003} Inaba, S., \& Ikoma, M.\ 2003, \aap, 410, 711

\bibitem[2003]{inaba_2003-2} Inaba, S., Wetherill, G.~W., \& Ikoma, M.\ 2003, \icarus, 166, 46

\bibitem[2009]{kley_2009} Kley, W., Bitsch, B., \& Klahr, H.\ 2009, \aap, 506, 971 

\bibitem[2000]{ki2000} Kokubo, E., \& Ida, S.\  2000, Icarus, 143, 15

\bibitem[2007]{johansen_2007} Johansen, A., Oishi, J.~S., Mac Low, M.-M., Klahr,
H., Henning, T., \& Youdin, A.\ 2007, \nat, 448, 1022 

\bibitem[2010]{li_2010} Li, S.~L., Agnor, C.~B., \& Lin, D.~N.~C.\ 2010, \apj,
720, 1161

\bibitem[2009]{lodders_2009} Lodders, K., Palme, H., \& Gail, H.~-.\ 2009,
Landolt-Bornstein, New Series, Astronomy and Astrophysics, Springer Verlag,
Berlin (print version, arXiv:0901.1149) 

\bibitem[2010]{machida_2010} Machida, M.~N., Kokubo, 
E., Inutsuka, S.-I., \& Matsumoto, T.\ 2010, \mnras, 405, 1227 

 \bibitem[2001]{masset_2001} Masset, F., \& Snellgrove, M.\ 2001, \mnras, 320,
L55 

\bibitem[2011]{miguel_2010-a} Miguel, Y., Guilera, O.~M., \& Brunini, A.\ 2011, \mnras, 412, 2113 

\bibitem[2010]{miguel_2010-b} Miguel, Y., Guilera, O.~M., \& Brunini, A.\ 2010, ``The Diversity of Planetary Systems Architectures:
Contrasting Theory with Observations'', submmited for publication in \mnras 

\bibitem[2008]{militzer} Militzer, B., Hubbard, W.~B., Vorberger, J., Tamblyn,
I., \& Bonev, S.~A.\ 2008, \apjl, 688, L45

\bibitem[2005]{morbi_2005} Morbidelli, A., 
Levison, H.~F., Tsiganis, K., \& Gomes, R.\ 2005, \nat, 435, 462 

\bibitem[2007]{morbi_2007} Morbidelli, A., \& Crida, A.\ 2007, \icarus, 191, 158

\bibitem[2007]{morbi_2007-2} Morbidelli, A., Tsiganis, K., Crida, A., Levison, H.~F., \& Gomes, R.\ 2007, \aj, 134, 1790

\bibitem[2009]{morbi_2009} Morbidelli, A., Brasser, R., Tsiganis, K., Gomes, R.,
\& Levison, H.~F.\ 2009, \aap, 507, 1041 

\bibitem[2009]{morbi_2009-2} Morbidelli, A., Bottke, W.~F., Nesvorn{\'y}, D., \&
Levison, H.~F.\ 2009, \icarus, 204, 558 

\bibitem[2010]{nelson_2010} Nelson, R.~P., \& Gressel, O.\ 2010, accepted for
publication in \mnras (print version, arXiv:1007.1144) 

\bibitem[2002]{ohtsuki} Ohtsuki, K., Stewart, G.~R., \& Ida, S.\ 2002, Icarus, 155, 436 

\bibitem[2010]{ormel} Ormel, C.~W., Dullemond, C.~P., \& Spaans, M.\ 2010,
\apjl, 714, L103

\bibitem[2010]{paardekooper_2010} Paardekooper, S.-J., Baruteau, C., Crida, A., \& Kley, W.\ 2010, \mnras, 401, 1950 

\bibitem[2000]{podol_2000} Podolak, M., Podolak, J.~I., \& Marley, M.~S.\ 2000,
\planss, 48, 143 

\bibitem[1996]{pollack_1996} Pollack, J.~B., Hubickyj, O., Bodenheimer, P., Lissauer, J.~J., Podolak, M., \& Greenzweig, Y.\ 1996, \icarus, 124, 62 

\bibitem[2002]{tanaka_2002} Tanaka, H., Takeuchi, T., \& Ward, W.~R.\ 2002, \apj, 565, 1257 
 
\bibitem[2003]{thommes} Thommes, E.~W., Duncan,  M.~J., \& Levison, H.~F.\ 2003,
Icarus, 161, 431 

\bibitem[2008]{thommes_2008} Thommes, E.~W., 
Matsumura, S., \& Rasio, F.~A.\ 2008, Science, 321, 814 

\bibitem[2005]{tsiga} Tsiganis, K., Gomes, R., Morbidelli, A., \& Levison,
H.~F.\ 2005, \nat, 435, 459 

\bibitem[1977]{weiden_1977} Weidenschilling, S.~J.\ 1977, \apss, 51, 153 

\bibitem[1997]{weiden_1997} Weidenschilling, S.~J.\ 1997, \icarus, 127, 290 

\bibitem[201)]{weiden_2010} Weidenschilling, S.~J.\ 2010, Lunar and Planetary Institute Science Conference Abstracts, 41, 1453 

\bibitem[1993]{ws} Wetherill, G.~W., \& Stewart, G.~R.\ 1993, Icarus, 106, 190

\bibitem[2011]{youdin_2011} Youdin, A.~N.\ 2011, arXiv:1102.4620 



\end{thebibliography}
\end{document}